\documentclass[stslayout, noinfoline]{imsart}
\usepackage{geometry}
\usepackage{graphicx}
\usepackage{amssymb, amsmath}
\usepackage{natbib}
\usepackage{epstopdf}
\usepackage{subfigure}
\usepackage{algorithm}
\usepackage[noend]{algpseudocode}
\usepackage{tcolorbox}
\usepackage{tikz}
\usepackage[colorlinks,citecolor=blue,urlcolor=blue]{hyperref}

\usetikzlibrary{intersections, backgrounds, calc}

\definecolor{light}{RGB}{220, 188, 188}
\definecolor{mid}{RGB}{185, 124, 124}
\definecolor{dark}{RGB}{143, 39, 39}
\definecolor{highlight}{RGB}{0, 255, 0}
\definecolor{gray10}{gray}{0.1}
\definecolor{gray20}{gray}{0.2}
\definecolor{gray30}{gray}{0.3}
\definecolor{gray40}{gray}{0.4}
\definecolor{gray60}{gray}{0.6}
\definecolor{gray70}{gray}{0.7}
\definecolor{gray80}{gray}{0.8}
\definecolor{gray90}{gray}{0.9}
\definecolor{gray95}{gray}{0.95}
\definecolor{comment}{gray}{0.50}

\graphicspath{{figures/}}

\begin{document}

\begin{frontmatter}

\title{Incomplete Reparameterizations\\
       and Equivalent Metrics}
\runtitle{Geometry of Incomplete Reparameterizations}

\begin{aug}
  \author{Michael Betancourt%
  \ead[label=e1]{betanalpha@gmail.com}}
  
  \runauthor{Betancourt}

  \address{Michael Betancourt is the principal research scientist
           at Symplectomorphic, LLC. \printead{e1}.}

\end{aug}

\begin{abstract}
Reparameterizing a probabilisitic system is common advice for improving 
the performance of a statistical algorithm like Markov chain Monte Carlo, 
even though in theory such reparameterizations should leave the system, 
and the performance of any algorithm, invariant.  In this paper I
show how the reparameterizations common in practice are only
incomplete reparameterizations which result in different interactions
between a target probabilistic system and a given algorithm.  I then 
consider how these changing interactions manifest in the context of 
Markov chain Monte Carlo algorithms defined on Riemannian manifolds.  
In particular I show how any incomplete reparameterization is equivalent 
to modifying the metric geometry directly.
\end{abstract}

\end{frontmatter}

\newpage

\setcounter{tocdepth}{3}
\tableofcontents

\newpage

In practice probabilistic systems are implemented in a specific
parameterization of the target space, with the target probability
distribution specified by its representative probability density
function.  Reparameterizing the target space modifies this
probability density function but not the probability distribution
that it represents.  In other words any probabilistic computation 
should yield equivalent results no matter the parameterization used.

That said, reparameterizations are known to alter the performance 
of algorithms that implement those probabilistic computations,
suggesting that the \emph{interaction} between the target probability
distribution and the algorithm is not invariant.  These interactions,
and how they relate to a given parameterization, are often opaque and 
difficult to understand.  Developing explicit criteria to identify 
which parameterization yields the highest performance for a given
circumstance is particularly challenging. 

The situation improves when the target space is a Riemannian 
manifold and the algorithm in question exploits that Riemannian 
structure, as is common for Markov chain Monte Carlo methods. Here
we can construct and then analyze a comprehensive geometry that 
encompasses both the target probabilistic system and the algorithmic 
system.  In particular a geometric analysis reveals that the
reparameterizations employed in practice are only \emph{incomplete}
reparameterizations, modifying the target geometry but not the
algorithmic geometry and hence changing the relationship between
the two.

In this paper I formalize the effect of incomplete reparameterizations
for Markov chain Monte Carlo algorithms defined on Riemannian 
manifolds and construct an implicit criterion for the optimal
reparameterization for a given target distribution.  I begin by 
reviewing the basics of Riemannian geometry, and Markov transitions
that exploit Riemannian geometry, before demonstrating the duality 
between reparameterizations of the target space and equivalent metric
geometries and introducing a heuristic criterion to identify optimal 
reparameterizations.  Finally I apply these results to latent Gaussian 
models and their common centered and non-centered parameterizations.

\section{Riemannian Manifolds}

For the rest of this paper I will assume familiarity with the basics
of differential geometry.  Part I of \cite{BaezEtAl:1994} provides 
an accessible introduction with \cite{Lee:2013} giving a more thorough 
reference of the concepts and notation that I will use here.  

Let our target distribution be defined on a $D$-dimensional smooth 
manifold, $Q$, with local coordinate functions denoted
$\{q^{1}(q), \ldots, q^{D}(q)\}$.

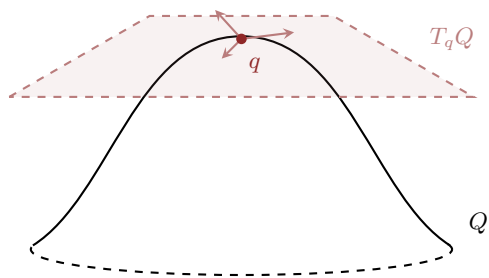
\begin{figure*}
\centering
\begin{tikzpicture}[scale=0.35, thick,
declare function={ fr(\t) = 3 * (1 - \t) * (1 - \t) * \t * 4
                           + 3 * (1 - \t) * \t * \t * 5 + \t * \t * \t * 8;},
declare function={ fh(\t) =  3 * (1 - \t) * \t * \t * 6 + \t * \t * \t * 8;} 
]
  \draw[dashed] (0, -8) circle [x radius=8, y radius={8 * sin(7)}]; 

  \filldraw[draw=black, fill=white] (-7.92, -7.85) .. controls (-5, -6) and (-4, 0.1) 
    .. (0, 0.1) .. controls (4, 0.1) and (5, -6) .. (7.92, -7.85);
    
  \node[] at (9, -7) { $Q$ };
  
  \pgfmathsetmacro{\dx}{0.4}
  \pgfmathsetmacro{\dy}{-0.075}

  \pgfmathsetmacro{\xo}{0}
  \pgfmathsetmacro{\yo}{0}
  \pgfmathsetmacro{\zo}{5}

  \pgfmathsetmacro{\bLx}{5}
  \pgfmathsetmacro{\bLy}{2.5}
  
  \pgfmathsetmacro{\Lx}{5}
  \pgfmathsetmacro{\Ly}{5}
  
  \pgfmathsetmacro{\dxo}{5}
  \pgfmathsetmacro{\dyo}{0.5}
  
  \pgfmathsetmacro{\phi}{60}

  \pgfmathsetmacro{\xnom}{\xo - \bLx}
  \pgfmathsetmacro{\ynom}{\yo - \bLy}
    
  \pgfmathsetmacro{\z}{\zo + sin(\phi) * \ynom}
  \pgfmathsetmacro{\bya}{cos(\phi) * \ynom * (\zo / \z)}
  \pgfmathsetmacro{\bxa}{\xnom * (\zo / \z)}    

  \pgfmathsetmacro{\xnom}{\xo - \bLx}
  \pgfmathsetmacro{\ynom}{\yo + \bLy}
    
  \pgfmathsetmacro{\z}{\zo + sin(\phi) * \ynom}
  \pgfmathsetmacro{\byb}{cos(\phi) * \ynom * (\zo / \z)}
  \pgfmathsetmacro{\bxb}{\xnom * (\zo / \z)}  
  
  \pgfmathsetmacro{\xnom}{\xo + \bLx}
  \pgfmathsetmacro{\ynom}{\yo + \bLy}
    
  \pgfmathsetmacro{\z}{\zo + sin(\phi) * \ynom}
  \pgfmathsetmacro{\byc}{cos(\phi) * \ynom * (\zo / \z)}
  \pgfmathsetmacro{\bxc}{\xnom * (\zo / \z)}      

  \pgfmathsetmacro{\xnom}{\xo + \bLx}
  \pgfmathsetmacro{\ynom}{\yo - \bLy}

  \pgfmathsetmacro{\z}{\zo + sin(\phi) * \ynom}
  \pgfmathsetmacro{\byd}{cos(\phi) * \ynom * (\zo / \z)}
  \pgfmathsetmacro{\bxd}{\xnom * (\zo / \z)} 
  
  \fill[mid, color=mid, opacity=0.10] (\bxa, \bya) -- (\bxb, \byb) -- (\bxc, \byc) -- (\bxd, \byd) -- cycle;
  \draw[mid, dashed] (\bxa, \bya) -- (\bxb, \byb) -- (\bxc, \byc) -- (\bxd, \byd) -- cycle;

  \node[color=mid] at (8, 0) { $T_{q} Q$ };

  \draw[color=mid, ->, >=stealth] (0, 0) -- (2, 0.25);
  \draw[color=mid, ->, >=stealth] (0, 0) -- (-0.75, -0.75);
  \draw[color=mid, ->, >=stealth] (0, 0) -- (-1, 1.1);
  
  \fill [fill=dark] (0, 0) circle (0.2);
  
  \node[color=dark] at (0.5, -1) { $q$ };

\end{tikzpicture}
\caption{
Each tangent space $T_{q} Q$ is a $D$-dimensional vector space associated 
with a point $q \in Q$.  If we embed the manifold $Q$ in a 
higher-dimensional space then we can interpret the tangent space as a 
plane fixed to $q$ and tangent to the manifold at that point of connection.
}
\label{fig:tangent_space} 
\end{figure*}

The tangent space at each point, $T_{q} Q$, is a vector space over the
real numbers whose elements can be associated with equivalence classes 
of one-dimensional curves sharing the same velocity at that point
(Figure \ref{fig:tangent_space}).  Local coordinate functions induce 
a basis within each tangent space given by the velocities of each
coordinate function, which we denote by the partial derivatives,
\begin{equation*}
\left\{ 
\frac{\partial}{\partial q^{1}}, \ldots, 
\frac{\partial}{\partial q^{D}}
\right\}
\equiv
\{ \partial_{1}, \ldots, \partial_{D} \}.
\end{equation*}

All of the tangent spaces in a manifold stitch together to define a 
$2D$-dimensional manifold with a canonical projection back down to
the base space, $\pi: TQ \rightarrow Q$, called the tangent bundle.
Vector fields are sections of this bundle, specifying a vector within
each tangent space, $v : Q \rightarrow TQ$.  The space of all vector
fields on $Q$ is denoted $\Gamma (Q)$.

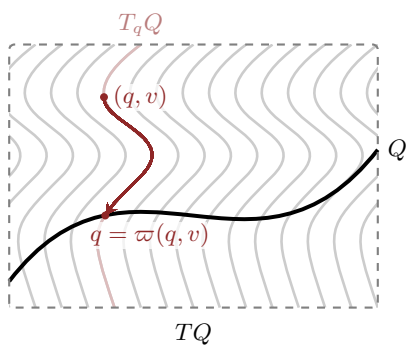
\begin{figure*}
\centering
\begin{tikzpicture}[scale=0.35, thick]
  \begin{scope}
    \clip (5, 0) rectangle +(14, 10);
    \foreach \i in {2, 3, ..., 22} {
      \draw [color=gray80, line width=1] (\i, 0) .. controls +(-1, 3) .. +(1, 5) .. 
                                         controls +(2, 2) and +(-4, -3) .. +(1, 10);
    }  
  \end{scope}
  
  \begin{scope}
    \clip (5, 0) rectangle +(14, 10);
      \draw [color=light, line width=1] (9, 0) .. controls +(-1, 3) .. +(1, 5) .. 
                                         controls +(2, 2) and +(-4, -3) .. +(1, 10); 
  \end{scope}
  
  \node[text=mid] at (10, 10.75) { $T_{q}Q$ };
  
  \draw [line width=1.5] (5, 1) .. controls (9.6, 7) and (14.3, 0) .. (19, 6)
  node[right] { $Q$ };
  
  \fill [fill=white, opacity=0.75] (9, 7.5) rectangle +(2, 1);
  \fill [fill=dark, text=dark] (8.6, 8) circle (0.15)
  node[right] { $(q, v)$ };

  \foreach \i in {2.5, 3, ..., 21.5} {
    \draw [color=dark, line width=1, <-, >=stealth] (8.65, 3.5) .. controls (9, 4) .. (10, 5) .. 
                                                    controls (11.5, 6.5) and (8.55, 7) .. (8.6, 8);

  }  
    
  \fill [fill=dark, text=dark] (8.65, 3.5) circle (0.15);
  \fill [fill=white, opacity=0.75] (8, 2.25) rectangle +(4.5, 1);
  \node[align=left, text=dark] at (10.5, 2.75) { $q = \varpi(q, v)$ };
  
  \draw [rounded corners=2pt, color=gray, dashed] (5, 0) rectangle +(14, 10);
  \node at (12, -1) { $TQ$ };
  
\end{tikzpicture}
\caption{
The tangent bundle $TQ$ is given by weaving together the tangent spaces 
attached at each point in the base manifold, $Q$.  Here the base manifold 
is one-dimensional and each tangent space can be represented with a 
one-dimensional line.  Each point in the tangent bundle is identified
by a point in the base space, $q \in Q$, and a point in the corresponding
tangent space, $v \in T_{q} Q$.  The tangent bundle is equipped with a
natural projection operator, $\varpi: TQ \rightarrow Q$ that maps each
point in the tangent space back to the associated point in the base space.
}
\label{fig:tangent_bundle} 
\end{figure*}

Similarly the cotangent space at each point, $T^{*}_{q} Q$ is a vector space
over the real numbers whose elements can be associated with equivalence 
classes of real-valued functions with the same first-order differential behavior.  
These covectors are also dual to vectors of the tangent space, with each covector 
mapping a vector to a real number and vice versa.  Within a local chart the 
coordinate functions define a basis for the cotangent space given by the gradients
of the coordinate functions,
\begin{equation*}
\{ \mathrm{d} q_{1}, \ldots, \mathrm{d} q_{D}\}.
\end{equation*}

As with the tangent spaces, all of the cotangent spaces can be weaved together 
to give the cotangent bundle, $\pi^{*}: T^{*}Q \rightarrow Q$.  Covector fields, 
or one-forms, are sections of this bundle, specifying a covector within each 
cotangent space, $\alpha : Q \rightarrow T^{*}Q$.

Without any additional structure a manifold $Q$ isn't particularly rigid; 
there is little structure within each tangent space, let alone between tangent
spaces.  In order to rigidify the manifold, and elevate it to a \emph{Riemannian 
manifold}, we need to equip it with additional structure.  In particular we 
need to specify a \emph{Riemannian metric} and a \emph{linear connection}
which allow us to compare vectors within a single tangent space as well as 
vectors in different tangent spaces.  Their structure also gives rise to 
\emph{geodesics} and the ability to flow through the manifold.

For a thorough introduction of Riemannian manifolds see \cite{Lee:2018}.  In
the next few sections I will review the basic concepts that we will need to
construct Markov transitions on Riemannian manifolds.

\subsection{Metrics} \label{sec:metrics}

A Riemannian metric a positive-defining map taking two vector fields to the 
real numbers,
\begin{alignat*}{6}
g :\; &\Gamma (Q) \times \Gamma (Q) & &\rightarrow& \; &\mathbb{R}&
\\
&(v, u)& &\mapsto& &g(u, v)&,
\end{alignat*}
such that $g(u, v) = g(v, u) > 0$ for any distinct $u, v \in \Gamma (Q)$
and $g(u, u) = 0$ only if $u = 0$.

\begin{figure*}
\centering
\begin{tikzpicture}[scale=0.35, thick,
declare function={ fr(\t) = 3 * (1 - \t) * (1 - \t) * \t * 4
                           + 3 * (1 - \t) * \t * \t * 5 + \t * \t * \t * 8;},
declare function={ fh(\t) =  3 * (1 - \t) * \t * \t * 6 + \t * \t * \t * 8;} 
]
  \draw[dashed] (0, -8) circle [x radius=8, y radius={8 * sin(7)}]; 

  \filldraw[draw=black, fill=white] (-7.92, -7.85) .. controls (-5, -6) and (-4, 0.1) 
    .. (0, 0.1) .. controls (4, 0.1) and (5, -6) .. (7.92, -7.85);
    
  \node[] at (9, -7) { $Q$ };
  
  \pgfmathsetmacro{\xo}{0}
  \pgfmathsetmacro{\yo}{0}
  \pgfmathsetmacro{\zo}{5}

  \pgfmathsetmacro{\bLx}{5}
  \pgfmathsetmacro{\bLy}{2.5}
  
  \pgfmathsetmacro{\Lx}{5}
  \pgfmathsetmacro{\Ly}{5}
  
  \pgfmathsetmacro{\dxo}{5}
  \pgfmathsetmacro{\dyo}{0.5}
  
  \pgfmathsetmacro{\theta}{0}
  \pgfmathsetmacro{\phi}{60}

  \pgfmathsetmacro{\xnom}{\xo - \bLx}
  \pgfmathsetmacro{\ynom}{\yo - \bLy}
    
  \pgfmathsetmacro{\z}{\zo + sin(\phi) * \ynom}
  \pgfmathsetmacro{\bya}{cos(\phi) * \ynom * (\zo / \z)}
  \pgfmathsetmacro{\bxa}{\xnom * (\zo / \z)}    

  \pgfmathsetmacro{\xnom}{\xo - \bLx}
  \pgfmathsetmacro{\ynom}{\yo + \bLy}
    
  \pgfmathsetmacro{\z}{\zo + sin(\phi) * \ynom}
  \pgfmathsetmacro{\byb}{cos(\phi) * \ynom * (\zo / \z)}
  \pgfmathsetmacro{\bxb}{\xnom * (\zo / \z)}  
  
  \pgfmathsetmacro{\xnom}{\xo + \bLx}
  \pgfmathsetmacro{\ynom}{\yo + \bLy}
    
  \pgfmathsetmacro{\z}{\zo + sin(\phi) * \ynom}
  \pgfmathsetmacro{\byc}{cos(\phi) * \ynom * (\zo / \z)}
  \pgfmathsetmacro{\bxc}{\xnom * (\zo / \z)}      

  \pgfmathsetmacro{\xnom}{\xo + \bLx}
  \pgfmathsetmacro{\ynom}{\yo - \bLy}

  \pgfmathsetmacro{\z}{\zo + sin(\phi) * \ynom}
  \pgfmathsetmacro{\byd}{cos(\phi) * \ynom * (\zo / \z)}
  \pgfmathsetmacro{\bxd}{\xnom * (\zo / \z)} 
  
  \fill[mid, color=mid, opacity=0.10] (\bxa, \bya) -- (\bxb, \byb) -- (\bxc, \byc) -- (\bxd, \byd) -- cycle;

  \begin{scope} 
    \clip (\bxa, \bya) -- (\bxb, \byb) -- (\bxc, \byc) -- (\bxd, \byd) -- cycle;

    \foreach \n in {-4, -3, ..., 4} {
      \pgfmathsetmacro{\xanom}{\xo - \Lx}
      \pgfmathsetmacro{\yanom}{\yo + (\n / 5) * \Ly)}
    
      \pgfmathsetmacro{\xarot}{cos(\theta) * \xanom + sin(\theta) * \yanom}
      \pgfmathsetmacro{\yarot}{-sin(\theta) * \xanom + cos(\theta) * \yanom}
    
      \pgfmathsetmacro{\za}{\zo + sin(\phi) * \yarot}
      \pgfmathsetmacro{\ya}{cos(\phi) * \yarot * (\zo / \za)}
      \pgfmathsetmacro{\xa}{\xarot * (\zo / \za)}    

      \pgfmathsetmacro{\xbnom}{\xo + \Lx}
      \pgfmathsetmacro{\ybnom}{\yo + (\n / 5) * \Ly)}
    
      \pgfmathsetmacro{\xbrot}{cos(\theta) * \xbnom + sin(\theta) * \ybnom}
      \pgfmathsetmacro{\ybrot}{-sin(\theta) * \xbnom + cos(\theta) * \ybnom}
    
      \pgfmathsetmacro{\zb}{\zo + sin(\phi) * \ybrot}
      \pgfmathsetmacro{\yb}{cos(\phi) * \ybrot * (\zo / \zb)}
      \pgfmathsetmacro{\xb}{\xbrot * (\zo / \zb)}   
    
      \draw[gray70, line width=0.25] ({\xa}, {\ya}) -- ({\xb}, {\yb});
    }
  
    \foreach \n in {-4, -3, ..., 4} {
      \pgfmathsetmacro{\xanom}{\xo + (\n / 5) * \Lx}
      \pgfmathsetmacro{\yanom}{\yo - \Ly}
    
      \pgfmathsetmacro{\xarot}{cos(\theta) * \xanom + sin(\theta) * \yanom}
      \pgfmathsetmacro{\yarot}{-sin(\theta) * \xanom + cos(\theta) * \yanom}
    
      \pgfmathsetmacro{\za}{\zo + sin(\phi) * \yarot}
      \pgfmathsetmacro{\ya}{cos(\phi) * \yarot * (\zo / \za)}
      \pgfmathsetmacro{\xa}{\xarot * (\zo / \za)}      

      \pgfmathsetmacro{\xbnom}{\xo + (\n / 5) * \Lx}
      \pgfmathsetmacro{\ybnom}{\yo + \Ly}
    
      \pgfmathsetmacro{\xbrot}{cos(\theta) * \xbnom + sin(\theta) * \ybnom}
      \pgfmathsetmacro{\ybrot}{-sin(\theta) * \xbnom + cos(\theta) * \ybnom}
    
      \pgfmathsetmacro{\zb}{\zo + sin(\phi) * \ybrot}
      \pgfmathsetmacro{\yb}{cos(\phi) * \ybrot * (\zo / \zb)}
      \pgfmathsetmacro{\xb}{\xbrot * (\zo / \zb)}   
    
      \draw[gray70, line width=0.25] ({\xa}, {\ya}) -- ({\xb}, {\yb});
    }
  \end{scope}
  
  \draw[mid, dashed] (\bxa, \bya) -- (\bxb, \byb) -- (\bxc, \byc) -- (\bxd, \byd) -- cycle;
  
  \node[color=mid] at (8, 0) { $T_{q} Q$ };
  
  \draw[color=mid, ->, >=stealth] (0, 0) -- (2, 0);
  \draw[color=mid, ->, >=stealth] (0, 0) -- (0, 0.65);
  
  \fill [fill=dark] (0, 0) circle (0.2);
  
  \node[color=dark] at (0.5, -1) { $q$ };

\end{tikzpicture}
\caption{
A metric adds rigidity to each tangent space, defining amongst other 
things orthogonal bases of vectors that coordinate the vector spaces.
}
\label{fig:metric} 
\end{figure*}
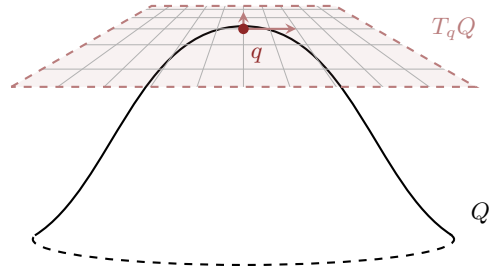

Within each tangent space the metric induces an inner product,
\begin{alignat*}{6}
g_{q} :\; &T_{q} Q \times T_{q} Q & &\rightarrow& \; &\mathbb{R}&
\\
&(v, u)& &\mapsto& &g_{q}(v, u)&,
\end{alignat*}
which allows us to orient vectors relative to each other.  In
particular the length of a vector is defined by 
\begin{equation*}
\vert\vert v \vert\vert = \sqrt{g_{q}(\vec{v}, \vec{v}) }
\end{equation*}
while the angle between two vectors is be defined by 
\begin{equation*}
\cos \theta = g_{q}(\vec{u}, \vec{v}).
\end{equation*}
If $g_{q}(\vec{u}, \vec{v}) = 0$ then the vectors are said to be 
perpendicular or orthogonal.  These concepts allow us to define, for 
example, orthonormal bases within a tangent space such that each basis 
vector has unit length and is orthogonal to each other basis vector 
(Figure \ref{fig:metric}).

One powerful feature of a metric is its ability to transform vector fields into
covector fields and vice versa.  In particular given a vector field $v$ we can 
define a corresponding covector field $v^{\flat}$ as the covector field satisfying
\begin{equation*}
v^{\flat} (u) = g(u, v),
\end{equation*}
for any vector field $u$.  The inverse of this map takes any covector field $\nu$ 
into a corresponding vector field, $\alpha^{\sharp}$.  This inverse transformation 
can then be used to define an \emph{inverse metric} over cotangent fields as
\begin{equation*}
g^{-1}(\alpha, \beta) = g(\alpha^{\sharp}, \beta^{\sharp}).
\end{equation*}  

By construction any metric is isomorphic to an element of the symmetric tensor 
product $T^{*} Q \otimes T^{*} Q$.  We can use this equivalence to represent a 
given metric in a local coordinate basis with the $D^{2}$ component functions
\begin{equation*}
g(q) = g_{ij}(q) \, \mathrm{d} q^{i} \otimes \mathrm{d} q^{j}.
\end{equation*}
The inverse metric is similarly locally represented by the component functions 
\begin{equation*}
g^{-1}(q) = g^{ij}(q) \ \partial_{i} \otimes \partial_{j},
\end{equation*}
where $g^{ij}(q)$ is the matrix inverse satisfying 
$g_{ij}(q) \cdot g^{jk}(q) = \delta^{k}_{i}$ for all $q \in Q$.

If a manifold admits an atlas such that the metric components equal the identity 
matrix, $g_{ij} = \delta_{ij}$, in \emph{every} chart then the manifold is said 
to be \emph{Euclidean}.  Unfortunately this definition is not how the term
"Euclidean" is used colloquially in statistics.  There "Euclidean" denotes 
any algorithm with constant metric component functions and "Riemannian" is used 
to denote a more general algorithm that exploits position-dependent metric 
component functions.  This distinction between component functions, however, is 
a property of the atlas being used, in particular the parameterizations within 
the local charts, and not the inherent structure of the manifold.  What makes a 
manifold Euclidean is not that its local metric component functions are constant 
but rather that they \emph{can be made constant with some choice of local 
parameterizations}.  For example any algorithm defined on the real numbers is 
geometrically a Euclidean algorithm \emph{no matter the parameterization used}.

\subsection{Connections}

A metric defines concepts like orientation and length which allows to compare
vectors \emph{within} each tangent space, but a manifold equipped with only a 
metric is still not rigid enough for us to compare vectors that live in 
\emph{different} tangent spaces.  To make the manifold fully rigid we need to 
introduce a \emph{connection} between these vector spaces.

We start differentially and ask whether or not we can compute directional 
derivatives of vector fields, in other words how vector fields change along 
a given direction, just as we can compute directional derivatives of functions.
Unlike directional derivatives of functions, however, there is no unique
directional derivatives of vector fields.  Instead we have to \emph{impose} 
one.

A linear connection defines a directional derivatives of vector fields through 
a \emph{covariant derivative} that maps two vector fields into a third,
\begin{alignat*}{6}
\nabla :\; &\Gamma (Q) \times \Gamma (Q) & &\rightarrow& \; &\Gamma(Q)&
\\
&(v, u)& &\mapsto& &\nabla_{v} u&.
\end{alignat*}
The first input defines directions at each point in the manifold along which 
changes will be probed, the second input defines the vector field being probed,
and the output defines the vector-valued changes in each tangent space.

In order for such a map to qualify as a derivative, however, it has to satisfy 
the usual properties of derivations.  It must, for example, be linear with 
respect to multiplying the probing directions by functions,
\begin{equation*}
\nabla_{f_{1} \cdot v + f_{2} \cdot u} w
= f_{1} \nabla_{v} w + f_{2} \nabla_{u} w,
\end{equation*}
for any two real-valued functions $f_{1}: Q \rightarrow \mathbb{R}$ and 
$f_{2}: Q \rightarrow \mathbb{R}$.  Moreover it must be linear with respect 
to multiplying probed vector field by constants,
\begin{equation*}
\nabla_{v} (a \cdot u + b \cdot w)
= a \nabla_{v} u + b \nabla_{v} w,
\end{equation*}
for any two real-value constants $a, b \in \mathbb{R}$.  Finally it must 
satisfy the Leibnitz rule with respect to multiplying the probed vector 
field by functions,
\begin{equation*}
\nabla_{v} (f \cdot u)
= f \nabla_{v} u + v(f) \cdot u.
\end{equation*}

Within a local coordinate basis the action of the covariant derivative becomes
\begin{equation*}
\nabla_{v} u = 
\left( v^{i} \frac{\partial y^{k}}{\partial q^{i}}
 + \Gamma^{k}_{ij} v^{i} u^{j} \right) \partial_{k}.
\end{equation*}
In other words the linear connection is completely specified with $D^{3}$ component
functions denoted $\Gamma^{k}_{ij}(q)$. These \emph{Christoffel} coefficients are not 
the components of a tensor but rather transform in a much complex way as they encode 
second-order differential information.  Only with careful combinations, such as in 
the above equation, do the non-tensorial components cancel to leave a well-defined 
geometric object.

There are an infinite number of connections on a given manifold but there is a unique 
connection that is naturally compatible with a given Riemannian metric.  In local 
coordinates basis the Christoffel coefficients for such a Riemannian or 
\emph{Levi-Cevita connection} are given by the functions
\begin{equation*}
\Gamma^{k}_{ij}(q) = \frac{1}{2} g^{kl}(q)
\left( 
  \frac{ \partial g_{jl} }{ \partial q_{i} }(q)
+ \frac{ \partial g_{il} }{ \partial q_{j} }(q)
+ \frac{ \partial g_{ij} }{ \partial q_{l} }(q) \right).
\end{equation*}
When discussing Riemannian manifolds a natural connection is often assumed to 
complement a given metric and fully rigidify the manifold.  Here, too, we will 
assume the choice of a Levi-Cevita connection.

\subsection{Going Places} \label{sec:geodesics}

The differential structure imposed by a connection immediately relates neighboring 
tangent spaces.  It can also relate distant tangent spaces when we traverse 
special curves through the base manifold, $Q$.

Recall that a curve is a smooth map from an interval of the real numbers into our 
manifold,
\begin{alignat*}{6}
c :\; &I & &\rightarrow& \; &Q&
\\
&t& &\mapsto& &c(t)&.
\end{alignat*}
In particular the points on a curve and their corresponding tangent spaces define 
a subset of the tangent bundle, with restricted vector fields defined as sections 
of this subset.  When restricted to this subset the covariant derivative will 
define how restricted vector fields change along the curve.

One restricted vector field inherent to any curve is the velocity vector field,
\begin{equation*}
\dot{c}(t) \in T_{c(t)}Q, \, \forall t \in I.
\end{equation*}
When this restricted vector field is placed into both arguments of the 
covariant derivative, $\nabla_{\dot{c}(t)} \dot{c}(t)$, the output defines 
how velocities change along the curve.  In other words it defines the 
\emph{acceleration} along the curve with respect to the chosen connection.  

\begin{figure*}
\centering
\begin{tikzpicture}[scale=0.35, thick,
declare function={ fr(\t) = 3 * (1 - \t) * (1 - \t) * \t * 4
                           + 3 * (1 - \t) * \t * \t * 5 + \t * \t * \t * 8;},
declare function={ fh(\t) =  3 * (1 - \t) * \t * \t * 6 + \t * \t * \t * 8;},
declare function={ dr(\t) =  3 * (4 + \t * (-6 + 5 * \t));},
declare function={ dh(\t) =  36 * (1 - \t) * \t + 6 * \t * \t;},
declare function={ x(\t, \theta) = fr(\t) * cos(-\theta); },
declare function={ y(\t, \theta) = -fh(\t) - fr(\t) * sin(-\theta) * sin(7); },
declare function={ dx(\t, \theta) = dr(\t) * cos(-\theta); },
declare function={ dy(\t, \theta) = -dh(\t) - dr(\t) * sin(-\theta) * sin(7); }
]
  \draw[dashed, line width=1] (0, -8) circle [x radius=8, y radius={8 * sin(7)}]; 

  \filldraw[draw=black, line width=1, fill=white] (-7.92, -7.85) .. controls (-5, -6) and (-4, 0.1) 
                                    .. (0, 0.1) .. controls (4, 0.1) and (5, -6) .. (7.92, -7.85);
  
  \node[] at (9, -7) { $Q$ };

  \pgfmathsetmacro{\theta}{300}
  \draw[domain=0:1, smooth, samples=10, variable=\t, line width=1, color=dark] 
    plot ({x(\t, \theta)}, {y(\t, \theta)});
  
  \draw [color=mid, ->, >=stealth] 
          ({x(0, \theta)}, {y(0, \theta)})
      -- +({0.2 * dx(0, \theta)}, {0.2 * dy(0, \theta)});
  \fill [fill=dark] ({x(0, \theta)}, {y(0, \theta)}) circle (0.15);
  \node[align=left] at (-2.2, 1) { $q = \exp_{0 \cdot v}(q)$ };

  \fill [fill=dark] ({x(0.62, \theta)}, {y(0.62, \theta)}) circle (0.15);
  \node[align=left] at (-1, -5.25) { $q_{t} = \exp_{t \cdot v}(q)$ };

\end{tikzpicture}
\caption{
Each point in the base manifold, $q \in Q$, and vector in the corresponding
tangent space, $v \in T_{q} Q$--in other words each point in the tangent 
bundle--identifies a unique geodesic curve through $Q$.  Following this
curve for a given time defines an exponential map that transports the 
initial point through $Q$.
}
\label{fig:geodesic} 
\end{figure*}
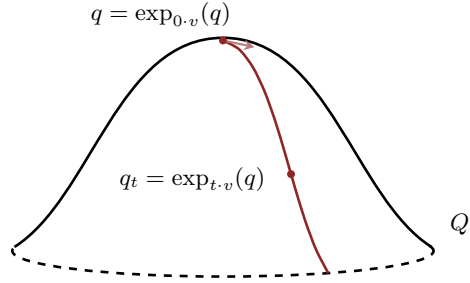

Curves with vanishing acceleration everywhere
\begin{equation*}
\nabla_{\dot{c}(t)} \dot{c}(t) = 0, \forall t \in I,
\end{equation*}
generalize the concept of a straight line to arbitrary smooth manifolds and 
are denoted \emph{geodesics}.  Geodesics have a variety of useful properties, 
but in the context of this paper one of the most useful is that they define 
a local flow on the base manifold.  Each point $q \in Q$ and vector 
$\vec{v} \in T_{q} Q$ intersects with only \emph{one} geodesic, defining an
unambiguous way to move through $Q$, at least within a local neighborhood 
where the geodesics are well-defined.  In other words once we pick a point 
and a direction we have a deterministic way to slide through the manifold 
(Figure \ref{fig:geodesic}).  The flow of the entire manifold along these 
geodesics is also known as the \emph{exponential map},
\begin{alignat*}{6}
\phi^{\exp} :\; &TQ \times \mathbb{R}& &\rightarrow& \; &Q&
\\
&(q, \vec{v}, t)& &\mapsto& &\phi^{\exp}_{t, \vec{v}}(q)&.
\end{alignat*}

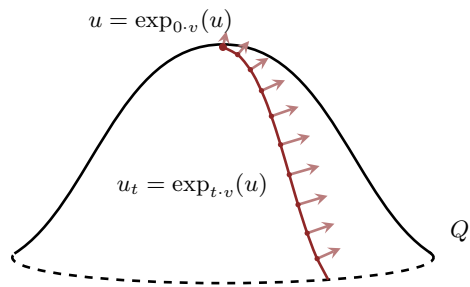
\begin{figure*}
\centering
\begin{tikzpicture}[scale=0.35, thick,
declare function={ fr(\t) = 3 * (1 - \t) * (1 - \t) * \t * 4
                           + 3 * (1 - \t) * \t * \t * 5 + \t * \t * \t * 8;},
declare function={ fh(\t) =  3 * (1 - \t) * \t * \t * 6 + \t * \t * \t * 8;},
declare function={ dr(\t) =  3 * (4 + \t * (-6 + 5 * \t));},
declare function={ dh(\t) =  36 * (1 - \t) * \t + 6 * \t * \t;},
declare function={ x(\t, \theta) = fr(\t) * cos(-\theta); },
declare function={ y(\t, \theta) = -fh(\t) - fr(\t) * sin(-\theta) * sin(7); },
declare function={ dx(\t, \theta) = dr(\t) * cos(-\theta); },
declare function={ dy(\t, \theta) = -dh(\t) - dr(\t) * sin(-\theta) * sin(7); }
]
  \draw[dashed, line width=1] (0, -8) circle [x radius=8, y radius={8 * sin(7)}]; 

  \filldraw[draw=black, line width=1, fill=white] (-7.92, -7.85) .. controls (-5, -6) and (-4, 0.1) 
                                    .. (0, 0.1) .. controls (4, 0.1) and (5, -6) .. (7.92, -7.85);
  
  \node[] at (9, -7) { $Q$ };

  \pgfmathsetmacro{\theta}{300}
  \draw[domain=0:1, smooth, samples=10, variable=\t, line width=1, color=dark] 
    plot ({x(\t, \theta)}, {y(\t, \theta)});

  \foreach \t in {0.9, 0.8, ..., 0} {
    \draw [color=mid, ->, >=stealth, line width=1] 
         ({x(\t, \theta)}, {y(\t, \theta)})
      -- +({-0.1 * dy(\t, \theta)}, {0.1 * dx(\t, \theta)});
    \fill [fill=dark] ({x(\t, \theta)}, {y(\t, \theta)}) circle (0.1);
  }
  
  \fill [fill=dark] ({x(0, \theta)}, {y(0, \theta)}) circle (0.15);
  \node[align=left] at (-2.2, 1) { $u = \exp_{0 \cdot v}(u)$ };

  \node[align=left] at (-1, -5.25) { $u_{t} = \exp_{t \cdot v}(u)$ };

\end{tikzpicture}
\caption{
A linear connection defines a transport of vectors in the tangent
space of any point along a geodesic to vectors in the tangent space
of any other point along that geodesic.  Here we transport a vector
$u \in T_{q}Q$ to the vector $u_{t} \in T_{q_{t}}Q$ where $q_{t}$ is
the exponential map of $(q, v) \in TQ$ for time $t$.}
\label{fig:parallel_transport} 
\end{figure*}

Geodesics, however, carry not only points along the manifold but also vectors 
from one tangent space to another.  Consider an initial point $q \in Q$, an 
initial direction $\vec{v} \in T_{q} Q$, and the corresponding geodesic curve 
with $c(t = 0) = q$.  For any vector $\vec{u} \in T_{q} Q$ there is a unique 
vector field restricted to the geodesic satisfying $u(c(0)) = \vec{u}$ and
$\nabla_{\dot{c}(t)} u = 0$ along the entire geodesic.  This restricted vector 
field defines \emph{parallel transport} of $\vec{u}$ along the geodesic
(Figure \ref{fig:parallel_transport}).  Overloading notation a bit I will also 
refer to this parallel transport as an exponential map,
\begin{alignat*}{6}
\phi^{\exp} :\; &Q \times T_{q} Q \times T_{q} Q \times \mathbb{R}& &\rightarrow& \; &\Gamma(Q)&
\\
&(q, \vec{v}, \vec{u}, t)& &\mapsto& &\phi^{\exp}_{t, \vec{v}}(\vec{u})&.
\end{alignat*}
By this definition the velocity vectors of a geodesic curve are all parallel 
transported into each other.

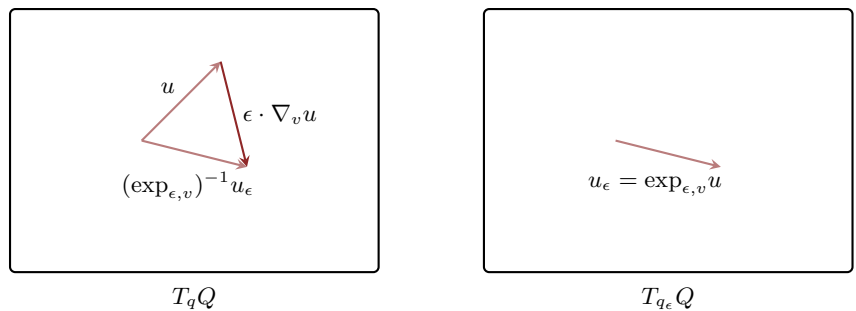
\begin{figure*}
\centering
\begin{tikzpicture}[scale=0.35, thick]
  \draw [rounded corners=2pt, color=black] (-2, 0) rectangle +(-14, 10);
  \node[] at (-9, -1) { $T_{q} Q$ };

  \draw [rounded corners=2pt, color=black] (2, 0) rectangle +(14, 10);
  \node[] at (9, -1) { $T_{q_{\epsilon}} Q$ };
  
  \node at (8.5, 3.25) { $u_{\epsilon} = \mathrm{exp}_{\epsilon, v} u$ };
  \draw [color=mid, ->, >=stealth] (7, 5) -- (11, 4);

  \node at (10 - 20, 7) { $u$ };
  \draw [color=mid, ->, >=stealth] (9 - 20, 5) -- (12 - 20, 8);

  \node at (10.75 - 20, 3.25) { $(\mathrm{exp}_{\epsilon, v})^{-1} u_{\epsilon}$ };
  \draw [color=mid, ->, >=stealth] (9 - 20, 5) -- (13 - 20, 4);

  \node at (14.25 - 20, 6) { $\epsilon \cdot \nabla_{v} u$ };
  \draw [color=dark, ->, >=stealth] (12 - 20, 8) -- (13 - 20, 4);
    
\end{tikzpicture}
\caption{
The covariant derivative can be interpreted as the difference between
a vector and its parallel transport for an infinitesimal amount of
time.  Here the initial point $q \in Q$ and initial vector $v \in T_{q} Q$
define a geodesic curve and the exponential map $q_{t}$ along that curve.
The vector $u \in T_{q} Q$ is parallel transported along the geodesic to 
the vector $u_{\epsilon} \in T_{q_{\epsilon}} Q$; the covariant derivative 
is the scaled difference between $u$ and the pullback of that parallel
transport.}
\label{fig:covariant_derviative} 
\end{figure*}

Parallel transport allows us to formalize the intuition for how a linear 
connection actually connects the tangent spaces in our manifold.  The 
first input of the covariant derivative defines directions, and 
corresponding geodesics, along which we probe the given vector field.  
The output of the covariant derivative is given by the change in that
probed vector field after being parallel transported for
an infinitesimal amount of time (Figure \ref{fig:covariant_derviative}),
\begin{equation*}
\cdot \nabla_{v} u 
= \lim_{\epsilon \rightarrow 0} 
\frac{ (\phi^{\exp}_{\epsilon, v})^{-1} u_{\epsilon} - u }{ \epsilon }.
\end{equation*}

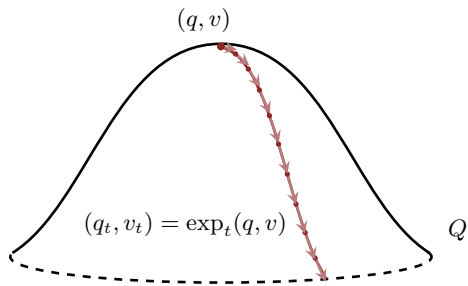
\begin{figure*}
\centering
\begin{tikzpicture}[scale=0.35, thick,
declare function={ fr(\t) = 3 * (1 - \t) * (1 - \t) * \t * 4
                           + 3 * (1 - \t) * \t * \t * 5 + \t * \t * \t * 8;},
declare function={ fh(\t) =  3 * (1 - \t) * \t * \t * 6 + \t * \t * \t * 8;},
declare function={ dr(\t) =  3 * (4 + \t * (-6 + 5 * \t));},
declare function={ dh(\t) =  36 * (1 - \t) * \t + 6 * \t * \t;},
declare function={ x(\t, \theta) = fr(\t) * cos(-\theta); },
declare function={ y(\t, \theta) = -fh(\t) - fr(\t) * sin(-\theta) * sin(7); },
declare function={ dx(\t, \theta) = dr(\t) * cos(-\theta); },
declare function={ dy(\t, \theta) = -dh(\t) - dr(\t) * sin(-\theta) * sin(7); }
]
  \draw[dashed, line width=1] (0, -8) circle [x radius=8, y radius={8 * sin(7)}]; 

  \filldraw[draw=black, line width=1, fill=white] (-7.92, -7.85) .. controls (-5, -6) and (-4, 0.1) 
                                    .. (0, 0.1) .. controls (4, 0.1) and (5, -6) .. (7.92, -7.85);
  
  \node[] at (9, -7) { $Q$ };
  \pgfmathsetmacro{\theta}{300}
  \draw[domain=0:1, smooth, samples=10, variable=\t, line width=1, color=dark] 
    plot ({x(\t, \theta)}, {y(\t, \theta)});

  \foreach \t in {0.9, 0.8, ..., 0.1} {
    \draw [color=mid, ->, >=stealth, line width=1] 
         ({x(\t, \theta)}, {y(\t, \theta)})
      -- +({0.1 * dx(\t, \theta)}, {0.1 * dy(\t, \theta)});
    \fill [fill=dark] ({x(\t, \theta)}, {y(\t, \theta)}) circle (0.1);
  }
  
  \draw [color=mid, ->, >=stealth, line width=1] 
          ({x(0, \theta)}, {y(0, \theta)})
      -- +({0.1 * dx(0, \theta)}, {0.1 * dy(0, \theta)});
  \fill [fill=dark] ({x(0, \theta)}, {y(0, \theta)}) circle (0.15);
  \node[align=left] at (-0.5, 1) { $(q, v)$ };

  \node[align=left] at (-1.2, -6.75) { $(q_{t}, v_{t}) = \exp_{t}(q, v)$ };
\end{tikzpicture}

\caption{
Any point $q \in Q$ and vector $v \in T_{q} Q$ defines a geodesic
which then transports both $q$ and $v$ along the curve.  Together
these transports defines a flow along the entire tangent bundle.
}
\label{fig:tangent_flow} 
\end{figure*}

Combining the geodesic flow and this parallel transport we see that the covariant 
derivative defines a flow along the entire tangent bundle 
(Figure \ref{fig:tangent_flow}).  An initial point $(q, \vec{v}) \in TQ$ defines a 
starting location and direction, which then identifies a unique geodesic path through 
$Q$.  At each point on that path we also have the velocity vectors of the geodesic 
which are the parallel transports of $\vec{v}$.  Overloading notation once again I 
will refer to this tangent flow as an exponential map,
\begin{alignat*}{6}
\phi^{\exp} :\; &TQ \times \mathbb{R}& &\rightarrow& \; &TQ&
\\
&(q, v, t)& &\mapsto& &\phi^{\exp}_{t}(q, v)&.
\end{alignat*}

\section{Riemannian Markov Transitions}

Markov chain Monte Carlo \citep{RobertEtAl:1999, BrooksEtAl:2011} explores 
a target probability distribution, $\pi(\mathrm{d}q)$, defined on $Q$ by 
sampling from a Markov transition conditioned on a given state, 
$\tau(\mathrm{d}q \mid q')$.  If the Markov transition preserves the target 
distribution in expectation,
\begin{equation*}
\pi(\mathrm{d}q) = \int \pi(\mathrm{d} q') \, \tau(\mathrm{d}q \mid q'),
\end{equation*}
then the repeated transitions generates a sequence of states that converges 
towards, and eventually disperses across, the typical the support of the target 
distribution.  The states in this Markov chain then define Markov chain Monte 
Carlo estimators
\begin{equation*}
\hat{f}_{N} 
= \frac{1}{N} \sum_{n = 1}^{N} f(q_{n})
\end{equation*}
that asymptotically converge to the true target expectation values,
\begin{equation*}
\lim_{N \rightarrow \infty} \hat{f}_{N} 
= \int \pi(\mathrm{d} q) \, f(q),
\end{equation*}
under typical regularity conditions.  The practical utility of a given Markov 
transition is determined by how quickly it explores target distribution and, 
consequently, how quickly the Markov chain Monte Carlo estimators converge to 
the true expectation values.

A powerful method for constructing Markov transitions is sampling over a family 
of deterministic transformations.  In particular, if $\phi_{t}$ is a family of 
continuous, bijective maps, $\phi_{t} : Q \rightarrow Q$, parameterized 
by $t \in T$, $\gamma$ is a probability distribution over $T$, and 
$\mathbb{I}_{A}(q)$ is the indicator function for the set $A \subset Q$, then
\begin{equation*}
\tau(\mathrm{d}q, q) 
= 
\int \gamma(\mathrm{d} t) 
\mathbb{I}_{\mathrm{d}q}(\phi_{t} (q)),
\end{equation*}
defines a Markov transition on $Q$ \citep{DiaconisEtAl:1999}.  If the 
transformations each preserve the target distribution,
\begin{equation*}
( (\phi_{t})_{*} \pi) (\mathrm{d}q) = \pi(\mathrm{d}q),
\end{equation*}
then this Markov transition will also preserve the target distribution and 
generate the desired Markov chains; when the individual transformations 
do not preserve the target distribution straightforward correction schemes 
can be applied to each move to ensure the desired invariance.  The freedom 
to choose a family of transformations and probability distribution over that 
family allows one to engineer particularly effective Markov transitions,
especially when those choices are informed by the structure of the target 
distribution itself.

Because Markov transitions condition on an initial state they can exploit 
the local structure of the target distribution within the neighborhood of 
that state to inform efficient transformations. In particular, if $Q$ is a 
Riemannian manifold then the local metric structure can be used to construct 
both families of deterministic transformations \emph{and} distributions 
over those families, defining potentially effective Markov transitions.  In 
this section we'll see how the local metric structure of a Riemannian 
manifold can be used to construct the ingredients of a Markov transition, 
and review examples of that construction that realize familiar algorithms.

\subsection{Geometric Transformations}

By exploiting the structure of the tangent and cotangent bundles associated 
with a manifold we can construct natural transformations that carry us around 
the space, providing the basis for Markov transitions.

\subsubsection{Tangent Flows}

As we saw in Section \ref{sec:geodesics}, equipping a smooth manifold with a 
Riemannian metric and its corresponding Levi-Cevita connection endows the 
space with natural geodesics that allow us to transport points and vectors 
along the curves.  These actions define a flow along the tangent bundle, $TQ$, 
which we referred to as an exponential map,
\begin{alignat*}{6}
\phi^{\exp} :\; &TQ \times \mathbb{R}& &\rightarrow& \; &TQ&
\\
&(q, \vec{v}, t)& &\mapsto& &\phi^{\exp}_{t}(q, \vec{v})&.
\end{alignat*}

Given an initial point we can identify a particular transformation by choosing 
a direction, which identifies a unique geodesic path, and a time, which informs 
how long to move along that path.  Dropping the final velocity vector then 
projects this cotangent flow to a flow across the base manifold, $Q$.  In 
other words the choice of vector and integration time parameterize a family of 
deterministic transformations on $Q$.

Because these transformations are not informed by the target distribution they 
will not, in general, preserve it.  Instead the tangent flow provides proposals 
that can be corrected to achieve the desired invariance.

\subsubsection{Cotangent Flows}

Unlike the tangent bundle, the cotangent bundle, $T^{*} Q$ is naturally equipped 
with a unique symplectic structure, $\omega$, and symplectic measure, $\Omega$, 
that allows to construct flows without the need for the extra structure 
introduced by a Riemannian metric \citep{BetancourtEtAl:2016a}.

Instead of complementing the manifold with a metric, we instead complement the 
cotangent bundle and its natural symplectic structure with some function
\begin{equation*}
H : T^{*} Q \rightarrow \mathbb{R},
\end{equation*}
denoted a Hamiltonian.  The choice of a Hamiltonian function immediately defines 
a Hamiltonian flow over the cotangent bundle, 
\begin{alignat*}{6}
\phi^{H} :\; &T^{*}Q \times \mathbb{R}& &\rightarrow& \; &T^{*}Q&
\\
&(q, p, t)& &\mapsto& &\phi^{H}_{t}(q, p)&.
\end{alignat*}

Hamiltonian flows have the added benefit of inherently preserving the canonical 
distribution, a probability distribution over the cotangent bundle given by
\begin{equation*}
\pi(\mathrm{d}q, \mathrm{d}p) = e^{-H(q, p)} \Omega(\mathrm{d}q, \mathrm{d}p).
\end{equation*}
If the Hamiltonian is chosen such that this canonical distribution projects to 
our target distribution, then the projection of the Hamiltonian flow will preserve
the target distribution, $\pi( \mathrm{d} q)$.  We can guarantee the desired 
invariance by introducing a conditional distribution over the cotangent fibers,
$\pi(\mathrm{d} p \mid q)$, defining the lifted joint distribution,
\begin{equation*}
\pi(\mathrm{d} q, \mathrm{d} p) 
= 
\pi(\mathrm{d} p \mid q) \, \pi( \mathrm{d} q),
\end{equation*}
and then taking the Hamiltonian to be the corresponding Radon-Nikodym derivative 
with respect to the symplectic measure,
\begin{equation*}
H 
= - \log 
\frac{ \mathrm{d} \pi (\mathrm{d}q, \mathrm{d}p) }
{ \mathrm{d} \Omega (\mathrm{d}q, \mathrm{d}p) }
= - \log \pi(p \mid q) - \log \pi(q).
\end{equation*}

Similar to the geodesic-informed tangent flow, this cotangent flow defines a 
family of transformations from any initial point $q \in Q$ parameterized by
the choice of cotangent vector, $p \in T^{*}_{q}Q$, which defines a unique 
Hamiltonian trajectory, and and integration time, $t$, which defines how long
to move along that trajectory.  Projecting this flow back to the base manifold 
defines the family of transformations from which we can construct a valid 
Markov transition.

Although these trajectories are not immediately dependent on a Riemannian metric, 
they do depend on the choice of Hamiltonian which itself depends on the choice of
some conditional probability distribution $\pi(\mathrm{d} p \mid q)$.  As we
will see in the next section, building such a conditional probability distribution 
is greatly facilitated by exploiting Riemannian metric structure.  Consequently 
in practice these cotangent flows are implicitly informed by the choice of
metric.

\subsection{Probability Distributions Over Moves}

Both the tangent and cotangent flows introduced above were parameterized by an 
integration time as well as an initial vector or covector.  In order to incorporate
these families of transformations into a Markov transition we need to impose 
probability distributions over these parameters.

Selecting a distribution over the real-valued integration times is straightforward, 
although it is not immediately obvious how to select an optimal distribution.  The 
choice of distribution of vectors and covectors, however, is complicated by the 
abstract geometry of the manifolds involved.  Fortunately the rigidity imposed by
a Riemannian metric drastically simplifies this problem.

Because the distribution of vectors and covectors can vary with the initial point 
we really want to define \emph{conditional} probability distributions over each 
of the tangent and cotangent spaces \citep{BetancourtEtAl:2016a}.  Conveniently 
a Riemannian metric provides all of the ingredients we need.  Within a given 
tangent space, for example the metric defines the quadratic form  $g_{q}(v, v)$ 
and the metric determinant, $| g(q) |$.  These are sufficient to construct any 
\emph{elliptical family of probability density functions} of the from
\begin{equation*}
\pi(v; q, \phi) = 
\xi ( g_{q}(v, v), \phi ) + \zeta ( | g(q) |, \phi ),
\end{equation*}
for appropriate choices of the real-value functions $\xi$ and $\zeta$.
Similarly in the cotangent spaces we can use the inverse metric to build 
elliptical probability density functions of the form
\begin{equation*}
\pi(p; q, \phi) = 
\iota ( g^{-1}_{q}(p, p), \phi ) + \kappa ( | g(q) |, \phi ).
\end{equation*}
Because the quadratic forms and metric determinants smoothly vary with the base 
point $q$, these probability density functions fuse together into well-defined
conditional probability density functions, $\pi(v \mid q; \phi)$ and 
$\pi(p \mid q; \phi)$.

Despite their relatively simple form elliptical families span a wide range of 
distributions, providing useful flexibility when constructing Markov transitions. 
For example elliptical families include not only the Gaussian family but also 
more heavy-tailed families like the Laplace and Cauchy families of probability
density functions.

\subsection{Example Constructions}

Metric-informed tangent flows and metric-informed tangent conditional distributions, 
provide the components of a full Markov transition.  We start at an initial point, 
$q$, and begin the transition by sampling an initial vector from the corresponding 
distribution over the tangent space, 
\begin{equation*}
v \sim \pi(\mathrm{d}v \mid q),
\end{equation*}
before sampling an integration time from some distribution that might be informed 
by this initial configuration,
\begin{equation*}
t \sim \pi(t \mid q, p).
\end{equation*}
The initial point and vector define a unique geodesic along which we integrate for 
time $t$, generating a random move to a new point in the tangent bundle which we 
can project back down to the base manifold.  The same construction holds for the 
cotangent bundle, using instead Hamiltonian trajectories and a metric-informed 
cotangent conditional distribution.

With some additional modifications to correct the transitions and preserve a 
desired target distribution, this general geometric procedure recovers quite a 
few well-known algorithms.  Here I demonstrate three -- random walk 
Metropolis-Hastings, Langevin Monte Carlo, and Hamiltonian Monte Carlo.

\subsubsection{Random Walk Metropolis-Hastings}

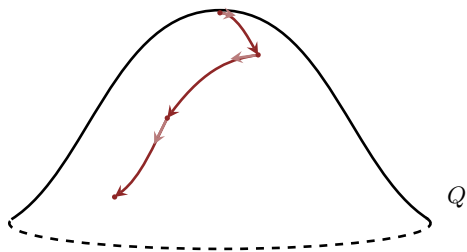
\begin{figure*}
\centering
\begin{tikzpicture}[scale=0.35, thick,
declare function={ fr(\t) = 3 * (1 - \t) * (1 - \t) * \t * 4
                           + 3 * (1 - \t) * \t * \t * 5 + \t * \t * \t * 8;},
declare function={ fh(\t) =  3 * (1 - \t) * \t * \t * 6 + \t * \t * \t * 8;},
declare function={ dr(\t) =  3 * (4 + \t * (-6 + 5 * \t));},
declare function={ dh(\t) =  36 * (1 - \t) * \t + 6 * \t * \t;},
declare function={ x(\t, \theta) = fr(\t) * cos(-\theta); },
declare function={ y(\t, \theta) = -fh(\t) - fr(\t) * sin(-\theta) * sin(7); },
declare function={ dx(\t, \theta) = dr(\t) * cos(-\theta); },
declare function={ dy(\t, \theta) = -dh(\t) - dr(\t) * sin(-\theta) * sin(7); }
]
  \draw[dashed, line width=1] (0, -8) circle [x radius=8, y radius={8 * sin(7)}]; 

  \filldraw[draw=black, line width=1, fill=white] (-7.92, -7.85) .. controls (-5, -6) and (-4, 0.1) 
                                    .. (0, 0.1) .. controls (4, 0.1) and (5, -6) .. (7.92, -7.85);

  \node[] at (9, -7) { $Q$ };

  \draw [color=dark, ->, >=stealth, line width=1] 
    (0, 0) .. controls (0.6, 0) and (1.1, -1) .. (1.45, -1.6);
  \draw [color=mid,  ->, >=stealth, line width=1] 
    (0, 0) -- +(0.6, -0.1);
  \fill [fill=dark] (0, 0) circle (0.1);

  \draw [color=dark, ->, >=stealth, line width=1] 
    (1.45, -1.6) .. controls (0.5, -1.75) and (-0.5, -2) .. (-2, -4);
  
  \draw [color=mid, ->, >=stealth, line width=1] 
         (1.45, -1.6) -- +(-1.05, -0.15);
  \fill [fill=dark] (1.45, -1.6) circle (0.1);

  \draw [color=dark, ->, >=stealth, line width=1] 
    (-2, -4) .. controls (-2.5, -5) and (-3, -6.25) .. (-4, -7);
  
  \draw [color=mid, ->, >=stealth, line width=1] 
         (-2, -4) -- +(-0.5, -1);
  \fill [fill=dark] (-2, -4) circle (0.1);
  
  \fill [fill=dark] (-4, -7) circle (0.1);
  
\end{tikzpicture}
\caption{
Randomly sampling a vector $v \in T_{q}Q$ and integration time, $t$, defines 
a transformation that takes an initial point $q$ to another point in the 
manifold.  Repeating this process defines a Markov chain over $Q$ whose
stationary distribution will depend on the choice of probability distributions
over the initial vectors and integration times.
}
\label{fig:random_walk} 
\end{figure*}

Repeatedly sampling a random direction and then following the corresponding 
geodesic for some finite time generates a second-order Markov process on the 
base manifold (Figure \ref{fig:random_walk}).  In the limit where the integration 
time vanishes this process converges to a random walk that diffuses across the 
manifold \citep{Hsu:2002}.

This diffusive behavior explores the manifold but it will not, in general, 
preserve a specified target distribution.  To achieve that behavior we need 
to correct the random walk, rejecting moves that stray too far from the typical
set of the joint distribution on the tangent bundle.

A standard approach to such corrections is to consider the moves as 
\emph{proposals} which are then accepted or rejected according to a 
Metropolis-Hastings correction.  The deterministic geodesic moves, however, 
require a small correction to serve as valid Metropolis proposals 
\citep{Tierney:1998}.  In order to admit a well-defined correction we have 
to compose each move with a reflection operator that flips the sign of the 
tangent vector after each flow,
\begin{align*}
R :& TQ \rightarrow TQ
\\
& (q, v) \mapsto (q, -v).
\end{align*}
This negation turns the flow into an \emph{involution} which returns to the 
initial state on the tangent bundle when the proposal is applied twice.

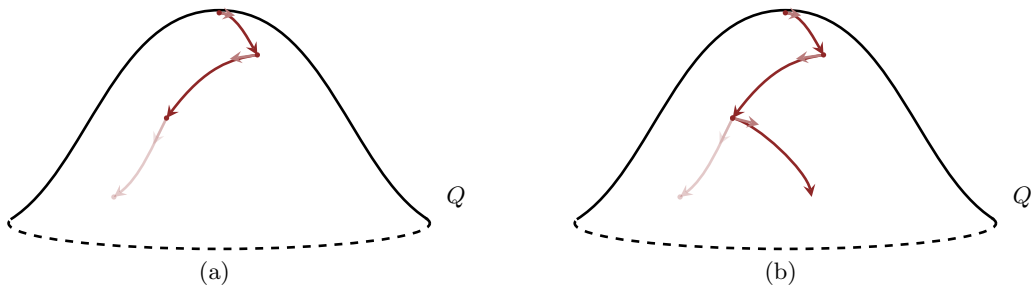
\begin{figure*}
\centering
\subfigure[]{
\begin{tikzpicture}[scale=0.35, thick,
declare function={ fr(\t) = 3 * (1 - \t) * (1 - \t) * \t * 4
                           + 3 * (1 - \t) * \t * \t * 5 + \t * \t * \t * 8;},
declare function={ fh(\t) =  3 * (1 - \t) * \t * \t * 6 + \t * \t * \t * 8;},
declare function={ dr(\t) =  3 * (4 + \t * (-6 + 5 * \t));},
declare function={ dh(\t) =  36 * (1 - \t) * \t + 6 * \t * \t;},
declare function={ x(\t, \theta) = fr(\t) * cos(-\theta); },
declare function={ y(\t, \theta) = -fh(\t) - fr(\t) * sin(-\theta) * sin(7); },
declare function={ dx(\t, \theta) = dr(\t) * cos(-\theta); },
declare function={ dy(\t, \theta) = -dh(\t) - dr(\t) * sin(-\theta) * sin(7); }
]
  \draw[color=white] (-10.25, 0) -- (10.25, 0);

  \draw[dashed, line width=1] (0, -8) circle [x radius=8, y radius={8 * sin(7)}]; 

  \filldraw[draw=black, line width=1, fill=white] (-7.92, -7.85) .. controls (-5, -6) and (-4, 0.1) 
                                    .. (0, 0.1) .. controls (4, 0.1) and (5, -6) .. (7.92, -7.85);
  
  \node[] at (9, -7) { $Q$ };

  \draw [color=dark, ->, >=stealth, line width=1] 
    (0, 0) .. controls (0.6, 0) and (1.1, -1) .. (1.45, -1.6);
  \draw [color=mid,  ->, >=stealth, line width=1] 
    (0, 0) -- +(0.6, -0.1);
  \fill [fill=dark] (0, 0) circle (0.1);

  \draw [color=dark, ->, >=stealth, line width=1] 
    (1.45, -1.6) .. controls (0.5, -1.75) and (-0.5, -2) .. (-2, -4);
  
  \draw [color=mid, ->, >=stealth, line width=1] 
         (1.45, -1.6) -- +(-1.05, -0.15);
  \fill [fill=dark] (1.45, -1.6) circle (0.1);

  \draw [color=dark, ->, >=stealth, line width=1, opacity=0.25] 
    (-2, -4) .. controls (-2.5, -5) and (-3, -6.25) .. (-4, -7);
  
  \draw [color=mid, ->, >=stealth, line width=1, opacity=0.25] 
         (-2, -4) -- +(-0.5, -1);
  \fill [fill=dark] (-2, -4) circle (0.1);
  
  \fill [fill=dark, opacity=0.25] (-4, -7) circle (0.1);
\end{tikzpicture}
}
\subfigure[]{
\begin{tikzpicture}[scale=0.35, thick,
declare function={ fr(\t) = 3 * (1 - \t) * (1 - \t) * \t * 4
                           + 3 * (1 - \t) * \t * \t * 5 + \t * \t * \t * 8;},
declare function={ fh(\t) =  3 * (1 - \t) * \t * \t * 6 + \t * \t * \t * 8;},
declare function={ dr(\t) =  3 * (4 + \t * (-6 + 5 * \t));},
declare function={ dh(\t) =  36 * (1 - \t) * \t + 6 * \t * \t;},
declare function={ x(\t, \theta) = fr(\t) * cos(-\theta); },
declare function={ y(\t, \theta) = -fh(\t) - fr(\t) * sin(-\theta) * sin(7); },
declare function={ dx(\t, \theta) = dr(\t) * cos(-\theta); },
declare function={ dy(\t, \theta) = -dh(\t) - dr(\t) * sin(-\theta) * sin(7); }
]
  \draw[color=white] (-10.25, 0) -- (10.25, 0);
  
  \draw[dashed, line width=1] (0, -8) circle [x radius=8, y radius={8 * sin(7)}]; 

  \filldraw[draw=black, line width=1, fill=white] (-7.92, -7.85) .. controls (-5, -6) and (-4, 0.1) 
                                    .. (0, 0.1) .. controls (4, 0.1) and (5, -6) .. (7.92, -7.85);
  
  \node[] at (9, -7) { $Q$ };

  \draw [color=dark, ->, >=stealth, line width=1] 
    (0, 0) .. controls (0.6, 0) and (1.1, -1) .. (1.45, -1.6);
  \draw [color=mid,  ->, >=stealth, line width=1] 
    (0, 0) -- +(0.6, -0.1);
  \fill [fill=dark] (0, 0) circle (0.1);

  \draw [color=dark, ->, >=stealth, line width=1] 
    (1.45, -1.6) .. controls (0.5, -1.75) and (-0.5, -2) .. (-2, -4);
  
  \draw [color=mid, ->, >=stealth, line width=1] 
         (1.45, -1.6) -- +(-1.05, -0.15);
  \fill [fill=dark] (1.45, -1.6) circle (0.1);

  \draw [color=dark, ->, >=stealth, line width=1, opacity=0.25] 
    (-2, -4) .. controls (-2.5, -5) and (-3, -6.25) .. (-4, -7);
  
  \draw [color=mid, ->, >=stealth, line width=1, opacity=0.25] 
         (-2, -4) -- +(-0.5, -1);
  \fill [fill=dark] (-2, -4) circle (0.1);
  
  \fill [fill=dark, opacity=0.25] (-4, -7) circle (0.1);

  \draw [color=dark, ->, >=stealth, line width=1] 
    (-2, -4) .. controls (-1, -4.25) and (0.8, -6) .. (1, -7);
  
  \draw [color=mid, ->, >=stealth, line width=1] 
         (-2, -4) -- +(1, -0.25);
  \fill [fill=dark] (-2, -4) circle (0.1);
  
\end{tikzpicture}
}
\caption{
(a) If a proposal strays too far from neighborhoods of high target
probability then a Metropolis correction is likely to reject that
proposal and return to the initial point.  (b) A proposal staying 
closer to high probability neighborhoods, however, will be accepted 
and ensure exploration that preserves the target distribution.
}
\label{fig:metropolis_correction} 
\end{figure*}

After sampling an initial velocity and time, flowing along the corresponding 
geodesic for that time, and then negating the final velocity we have a valid
proposal that can be accepted only with probability
\begin{equation*}
\mathbb{P}[\text{accept}] = \min( 1, r(q, v) ),
\end{equation*}
otherwise returning to the initial state ready for another transition 
(Figure \ref{fig:metropolis_correction}).  Here $r$ is the Radon-Nikodym 
derivative between the joint distribution on the tangent bundle and its 
pullback under the tangent flow,
\begin{equation*}
r(q, v) = \frac{ \mathrm{d}  (\phi^{\exp}_{t})_{*} \pi }{ \mathrm{d} \pi} (q, v).
\end{equation*}
Because of the careful dependence of the acceptance probability on the 
joint distribution, the complete transition will always preserves 
the joint distribution.  The marginal chain over the base manifold will
then preserve the target distribution.

In local coordinates on the tangent bundle the Radon-Nikodym derivative 
becomes a ratio of joint probability density functions,
\begin{align*}
r(q, v) 
&= 
\frac{ \pi(q') \, \pi(-v' \mid q' ) }
{ \pi(q) \, \pi(v \mid q) },
\end{align*}
where
\begin{equation*}
(q', v') = \phi^{\exp}_{t} (q, v).
\end{equation*}
The ratio of target probability density functions is known as the Metropolis 
ratio, while the ratio of tangent conditional probability density functions
is known as the Hastings ratio.

In the global coordinates of a Euclidean manifold this construction reduces 
to the usual random walk Metropolis algorithm.  If we specify the tangent 
distribution with a multivariate Gaussian probability density function then 
the process of sampling a tangent vector and flowing exactly yields a sample 
from a multivariate Gaussian on the base manifold whose covariance matrix is 
given by the components of the metric scaled by the integration time,
\begin{equation*}
\Sigma_{ij}(q) = t \cdot g_{ij}(q),
\end{equation*}
as the algorithm is typically presented.

For infinitesimally small integration times the geodesic random walk without 
any Metropolis correction defines a Brownian motion over the base manifold.
Taking small, but finite, integration times then provides a discrete approximation 
to that Brownian motion.  Introducing the Metropolis correction guides the 
discretized random walk towards the neighborhoods of high target probability, 
albeit relatively inefficiently in most contemporary problems.

\subsubsection{Langevin Monte Carlo}

A similar procedure applies to the cotangent bundle.  Sampling a covector from a 
cotangent conditional distribution and then applying the Hamiltonian flow for some
time defines a second-order stochastic process across the cotangent bundle.  Unlike 
the geodesic random walk, however, the Hamiltonian-informed process manifestly 
preserves the joint distribution on the cotangent bundle, and hence the marginal 
process preserves the target distribution on the base manifold.  For infinitesimally 
small integration times this process defines an Ornstein-Uhlenbeck over the base
manifold, a drifting diffusion that exactly targets the given joint distribution.

Unfortunately this invariance isn't robust enough to manifest exactly in practical 
applications where we have to \emph{approximate} the Hamiltonian flow with the 
discrete trajectories of a symplectic integrator, $\Phi^{H}_{\epsilon, L}$.  Here 
$\epsilon$ denotes the step size of the integration and $L$ the number of steps.  
Approximating the infinitesimal action of the Hamiltonian flow with one step of a 
symplectic integrator, $\Phi^{H}_{\epsilon, 1}$ gives \emph{Langevin Monte Carlo}, 
or sometimes \emph{unadjusted Langevin Monte Carlo} \citep{XifaraEtAl:2014}.

Although symplectic integrators are exceptionally accurate they are not perfect, and 
the numerical errors they introduce will bias the discrete transitions away from the 
target distribution.  In order to preserve the invariance of the target distribution,
especially in higher dimensions, we need to apply a Metropolis correction just as we 
did for the geodesics.  As in that case we first turn the discrete update into an 
involution with the composition of a reflection operator,
\begin{align*}
R :& T^{*}Q \rightarrow T^{*}Q
\\
& (q, p) \mapsto (q, -p),
\end{align*}
before applying a Metropolis correction that accepts the updated state only with 
probability
\begin{equation*}
\mathbb{P}[\text{accept}] = \min( 1, r(q, v) ),
\end{equation*}
where $r$ is the Radon-Nikodym derivative 
\begin{equation*}
r(q, p) = \frac{ \mathrm{d}  (\Phi^{H}_{\epsilon, 1})_{*} \pi }{ \mathrm{d} \pi} (q, p).
\end{equation*}
Combining the discrete Langevin dynamics with a Metropolis correction defines 
\emph{adjusted Langevin Monte Carlo}, or \emph{Metropolis adjusted Langevin
Monte Carlo}, or typically just \emph{MALA}.

Because the Metropolis correction is compensating only for the errors introduced by 
the symplectic integrator, and not the imperceptive geodesics of random walk
Metropolis-Hastings, Metropolis adjusted Langevin methods perform much better than 
their random walk equivalents.  Still, their overall performance is limited by the
diffusive nature of the transitions.

\subsubsection{Hamiltonian Monte Carlo}

To fully exploit Hamiltonian flow we need to follow it for much longer than 
infinitesimal times, taking advantage of the coherent trajectories to rapidly explore 
the target distribution.  In practice this is accomplished by first sampling a 
covector from the cotangent conditional distribution and then applying a symplectic 
integrator for multiple steps, $\Phi^{H}_{\epsilon, L}$, to simulate the Hamiltonian 
flow for time $t = \epsilon \cdot L$.  This defines the family of 
\emph{Hamiltonian Monte Carlo} methods \citep{BetancourtEtAl:2016a}.

Once we have a longer discrete trajectory we still have to correct for the small but 
non-negligible numerical errors.  The first Hamiltonian Monte Carlo methods considered 
only the final state in the trajectory, applying a Metropolis correction to that state 
as with random walk Metropolis and Metropolis adjusted Langevin methods.  Modern 
implementations, however, take advantage of the entire trajectory by going beyond 
Metropolis corrections.  For a thorough discussion see \citep{Betancourt:2018a}.

\section{Reparameterizations and Equivalent Metrics}

One of the benefits of the pure geometric construction that we have so far discussed 
is that it explicitly guides implementations.  Once local coordinates have been chosen,
the probability distributions, namely the target distribution and the tangent or cotangent
conditional distributions, can be specified with local probability density functions.
Likewise any geometric objects, namely the metric, can be specified with local component 
functions.  In this way everything is manifestly compatible with each other and the chosen
coordinates.  Under a reparameterization we simply begin with a new coordinate system and 
start the process anew.

That progression from geometric to coordinate, however, is not how algorithms are typically 
implemented in practice.  Instead practitioners often begin with a default local representation 
of the algorithmic system and are responsible for working out how that local representation 
transforms under reparameterizations.  These transformations are challenging to determine and
prone to error; cnsequently practitioners routinely neglect transforming the algorithmic
structure entirely, resulting in an \emph{incomplete} transformations and an entirely different 
Riemannian geometry.

In this section I first review the proper way local representations transform under a 
reparameterization of the local coordinates before considering the incomplete 
reparameterizations typical of practice.  In particular I explicitly derive the modified 
geometries that result from these incomplete reparameterizations.  Finally I use this 
relationship between incomplete reparameterizations and modified geometries to motivate an 
optimality criterion for reparameterizations in the context of a given target distribution.

\subsection{Complete Reparameterizations}

There are two equivalent perspectives on the reparameterization of a manifold: the 
\emph{passive} and the \emph{active} \cite{BaezEtAl:1994}.  In the passive perspective 
a reparameterization fixes the manifold but transforms each chart, while in the active 
perspective a reparameterization transforms the manifold while fixing the charts.  Although 
these two perspectives are equivalent the latter is closer to how reparameterizations are
implemented in practice and consequently I will focus on that perspective.

More formally in the active perspective a reparameterization is a diffeomorphism from 
the base manifold into itself,
\begin{alignat*}{6}
\psi :\; &Q& &\rightarrow& \; &Q&
\\
&q& &\mapsto& &q' = \psi(q)&.
\end{alignat*}
that pushes each chart forward.  These chart maps are linear transformations represented 
by \emph{Jacobian matrices},
\begin{equation*}
J^{i}_{j} (q) = \frac{ \partial \psi^{i} }{ \partial q^{j} } (q).
\end{equation*}
In other words even a non-linear reparameterization acts like a linear transformation 
within each local neighborhood.

For example, the coordinate functions in the new charts are given by
\begin{equation*}
(q')^{i}(q') = J^{i}_{j} (\psi^{-1}(q')) \cdot q^{j} (\psi^{-1}(q')).
\end{equation*}
Similarly the probability density function representation of a probability distribution 
within a given chart transforms by acquiring a factor of the inverse determinant of the 
Jacobian matrix,
\begin{equation*}
\pi(q') = \pi(\psi^{-1}(q')) 
\left| J (\psi^{-1}(q')) \right|^{-1}.
\end{equation*}

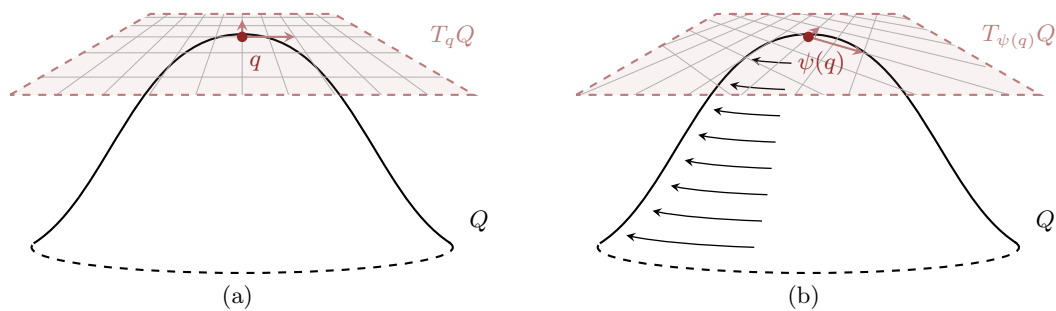
\begin{figure*}
\centering
\subfigure[]{
\begin{tikzpicture}[scale=0.35, thick,
declare function={ fr(\t) = 3 * (1 - \t) * (1 - \t) * \t * 4
                           + 3 * (1 - \t) * \t * \t * 5 + \t * \t * \t * 8;},
declare function={ fh(\t) =  3 * (1 - \t) * \t * \t * 6 + \t * \t * \t * 8;} 
]
  \draw[color=white] (-10.25, 0) -- (10.25, 0);
  
  \draw[dashed] (0, -8) circle [x radius=8, y radius={8 * sin(7)}]; 

  \filldraw[draw=black, fill=white] (-7.92, -7.85) .. controls (-5, -6) and (-4, 0.1) 
    .. (0, 0.1) .. controls (4, 0.1) and (5, -6) .. (7.92, -7.85);
  
  \node[] at (9, -7) { $Q$ };
    
  \pgfmathsetmacro{\xo}{0}
  \pgfmathsetmacro{\yo}{0}
  \pgfmathsetmacro{\zo}{5}

  \pgfmathsetmacro{\bLx}{5}
  \pgfmathsetmacro{\bLy}{2.5}
  
  \pgfmathsetmacro{\Lx}{5}
  \pgfmathsetmacro{\Ly}{5}
  
  \pgfmathsetmacro{\dxo}{5}
  \pgfmathsetmacro{\dyo}{0.5}
  
  \pgfmathsetmacro{\theta}{0}
  \pgfmathsetmacro{\phi}{60}

  \pgfmathsetmacro{\xnom}{\xo - \bLx}
  \pgfmathsetmacro{\ynom}{\yo - \bLy}
    
  \pgfmathsetmacro{\z}{\zo + sin(\phi) * \ynom}
  \pgfmathsetmacro{\bya}{cos(\phi) * \ynom * (\zo / \z)}
  \pgfmathsetmacro{\bxa}{\xnom * (\zo / \z)}    

  \pgfmathsetmacro{\xnom}{\xo - \bLx}
  \pgfmathsetmacro{\ynom}{\yo + \bLy}
    
  \pgfmathsetmacro{\z}{\zo + sin(\phi) * \ynom}
  \pgfmathsetmacro{\byb}{cos(\phi) * \ynom * (\zo / \z)}
  \pgfmathsetmacro{\bxb}{\xnom * (\zo / \z)}  
  
  \pgfmathsetmacro{\xnom}{\xo + \bLx}
  \pgfmathsetmacro{\ynom}{\yo + \bLy}
    
  \pgfmathsetmacro{\z}{\zo + sin(\phi) * \ynom}
  \pgfmathsetmacro{\byc}{cos(\phi) * \ynom * (\zo / \z)}
  \pgfmathsetmacro{\bxc}{\xnom * (\zo / \z)}      

  \pgfmathsetmacro{\xnom}{\xo + \bLx}
  \pgfmathsetmacro{\ynom}{\yo - \bLy}

  \pgfmathsetmacro{\z}{\zo + sin(\phi) * \ynom}
  \pgfmathsetmacro{\byd}{cos(\phi) * \ynom * (\zo / \z)}
  \pgfmathsetmacro{\bxd}{\xnom * (\zo / \z)} 
  
  \fill[mid, color=mid, opacity=0.10] (\bxa, \bya) -- (\bxb, \byb) -- (\bxc, \byc) -- (\bxd, \byd) -- cycle;

  \begin{scope} 
    \clip (\bxa, \bya) -- (\bxb, \byb) -- (\bxc, \byc) -- (\bxd, \byd) -- cycle;

    \foreach \n in {-4, -3, ..., 4} {
      \pgfmathsetmacro{\xanom}{\xo - \Lx}
      \pgfmathsetmacro{\yanom}{\yo + (\n / 5) * \Ly)}
    
      \pgfmathsetmacro{\xarot}{cos(\theta) * \xanom + sin(\theta) * \yanom}
      \pgfmathsetmacro{\yarot}{-sin(\theta) * \xanom + cos(\theta) * \yanom}
    
      \pgfmathsetmacro{\za}{\zo + sin(\phi) * \yarot}
      \pgfmathsetmacro{\ya}{cos(\phi) * \yarot * (\zo / \za)}
      \pgfmathsetmacro{\xa}{\xarot * (\zo / \za)}    

      \pgfmathsetmacro{\xbnom}{\xo + \Lx}
      \pgfmathsetmacro{\ybnom}{\yo + (\n / 5) * \Ly)}
    
      \pgfmathsetmacro{\xbrot}{cos(\theta) * \xbnom + sin(\theta) * \ybnom}
      \pgfmathsetmacro{\ybrot}{-sin(\theta) * \xbnom + cos(\theta) * \ybnom}
    
      \pgfmathsetmacro{\zb}{\zo + sin(\phi) * \ybrot}
      \pgfmathsetmacro{\yb}{cos(\phi) * \ybrot * (\zo / \zb)}
      \pgfmathsetmacro{\xb}{\xbrot * (\zo / \zb)}   
    
      \draw[gray70, line width=0.25] ({\xa}, {\ya}) -- ({\xb}, {\yb});
    }
  
    \foreach \n in {-4, -3, ..., 4} {
      \pgfmathsetmacro{\xanom}{\xo + (\n / 5) * \Lx}
      \pgfmathsetmacro{\yanom}{\yo - \Ly}
    
      \pgfmathsetmacro{\xarot}{cos(\theta) * \xanom + sin(\theta) * \yanom}
      \pgfmathsetmacro{\yarot}{-sin(\theta) * \xanom + cos(\theta) * \yanom}
    
      \pgfmathsetmacro{\za}{\zo + sin(\phi) * \yarot}
      \pgfmathsetmacro{\ya}{cos(\phi) * \yarot * (\zo / \za)}
      \pgfmathsetmacro{\xa}{\xarot * (\zo / \za)}      

      \pgfmathsetmacro{\xbnom}{\xo + (\n / 5) * \Lx}
      \pgfmathsetmacro{\ybnom}{\yo + \Ly}
    
      \pgfmathsetmacro{\xbrot}{cos(\theta) * \xbnom + sin(\theta) * \ybnom}
      \pgfmathsetmacro{\ybrot}{-sin(\theta) * \xbnom + cos(\theta) * \ybnom}
    
      \pgfmathsetmacro{\zb}{\zo + sin(\phi) * \ybrot}
      \pgfmathsetmacro{\yb}{cos(\phi) * \ybrot * (\zo / \zb)}
      \pgfmathsetmacro{\xb}{\xbrot * (\zo / \zb)}   
    
      \draw[gray70, line width=0.25] ({\xa}, {\ya}) -- ({\xb}, {\yb});
    }
  \end{scope}
  
  \draw[mid, dashed] (\bxa, \bya) -- (\bxb, \byb) -- (\bxc, \byc) -- (\bxd, \byd) -- cycle;
   \node[color=mid] at (8, 0) { $T_{q} Q$ };
  
  \draw[color=mid, ->, >=stealth] (0, 0) -- (2, 0);
  \draw[color=mid, ->, >=stealth] (0, 0) -- (0, 0.65);
  
  \fill [fill=dark] (0, 0) circle (0.2);
  \node[color=dark] at (0.5, -1) { $q$ };

\end{tikzpicture}
}
\subfigure[]{
\begin{tikzpicture}[scale=0.35, thick,
declare function={ fr(\t) = 3 * (1 - \t) * (1 - \t) * \t * 4
                           + 3 * (1 - \t) * \t * \t * 5 + \t * \t * \t * 8;},
declare function={ fh(\t) =  3 * (1 - \t) * \t * \t * 6 + \t * \t * \t * 8;} 
]
  \draw[color=white] (-10.25, 0) -- (10.25, 0);

  \draw[dashed] (0, -8) circle [x radius=8, y radius={8 * sin(7)}]; 

  \filldraw[draw=black, fill=white] (-7.92, -7.85) .. controls (-5, -6) and (-4, 0.1) 
    .. (0, 0.1) .. controls (4, 0.1) and (5, -6) .. (7.92, -7.85);
    
  \def\radii{{8, 6.85, 6.05, 5.435, 4.845, 4.175, 3.45, 2.5}}
  \def\heights{{8, 7, 6, 5, 4, 3, 2, 1}}
  \foreach \n in {0, 1, ..., 7} {
    \pgfmathparse{\radii[\n]}\edef\r{\pgfmathresult}
    \pgfmathparse{\heights[\n]}\edef\h{\pgfmathresult}
    \draw[line width=0.5, color=black, ->, >=stealth] ({0 - \r * cos(7) * sin(15)}, -\h) arc(255:210:{\r} and {\r * sin(7)}); 
  }
  
  \node[] at (9, -7) { $Q$ };
    
  \pgfmathsetmacro{\xo}{0}
  \pgfmathsetmacro{\yo}{0}
  \pgfmathsetmacro{\zo}{5}

  \pgfmathsetmacro{\bLx}{5}
  \pgfmathsetmacro{\bLy}{2.5}
  
  \pgfmathsetmacro{\Lx}{5}
  \pgfmathsetmacro{\Ly}{5}
  
  \pgfmathsetmacro{\dxo}{5}
  \pgfmathsetmacro{\dyo}{0.5}
  
  \pgfmathsetmacro{\theta}{30}
  \pgfmathsetmacro{\phi}{60}

  \pgfmathsetmacro{\xnom}{\xo - \bLx}
  \pgfmathsetmacro{\ynom}{\yo - \bLy}
    
  \pgfmathsetmacro{\z}{\zo + sin(\phi) * \ynom}
  \pgfmathsetmacro{\bya}{cos(\phi) * \ynom * (\zo / \z)}
  \pgfmathsetmacro{\bxa}{\xnom * (\zo / \z)}    

  \pgfmathsetmacro{\xnom}{\xo - \bLx}
  \pgfmathsetmacro{\ynom}{\yo + \bLy}
    
  \pgfmathsetmacro{\z}{\zo + sin(\phi) * \ynom}
  \pgfmathsetmacro{\byb}{cos(\phi) * \ynom * (\zo / \z)}
  \pgfmathsetmacro{\bxb}{\xnom * (\zo / \z)}  
  
  \pgfmathsetmacro{\xnom}{\xo + \bLx}
  \pgfmathsetmacro{\ynom}{\yo + \bLy}
    
  \pgfmathsetmacro{\z}{\zo + sin(\phi) * \ynom}
  \pgfmathsetmacro{\byc}{cos(\phi) * \ynom * (\zo / \z)}
  \pgfmathsetmacro{\bxc}{\xnom * (\zo / \z)}      

  \pgfmathsetmacro{\xnom}{\xo + \bLx}
  \pgfmathsetmacro{\ynom}{\yo - \bLy}

  \pgfmathsetmacro{\z}{\zo + sin(\phi) * \ynom}
  \pgfmathsetmacro{\byd}{cos(\phi) * \ynom * (\zo / \z)}
  \pgfmathsetmacro{\bxd}{\xnom * (\zo / \z)} 
  
  \fill[mid, color=mid, opacity=0.10] (\bxa, \bya) -- (\bxb, \byb) -- (\bxc, \byc) -- (\bxd, \byd) -- cycle;

  \begin{scope} 
    \clip (\bxa, \bya) -- (\bxb, \byb) -- (\bxc, \byc) -- (\bxd, \byd) -- cycle;

    \foreach \n in {-4, -3, ..., 4} {
      \pgfmathsetmacro{\xanom}{\xo - \Lx}
      \pgfmathsetmacro{\yanom}{\yo + (\n / 5) * \Ly)}
    
      \pgfmathsetmacro{\xarot}{cos(\theta) * \xanom + sin(\theta) * \yanom}
      \pgfmathsetmacro{\yarot}{-sin(\theta) * \xanom + cos(\theta) * \yanom}
    
      \pgfmathsetmacro{\za}{\zo + sin(\phi) * \yarot}
      \pgfmathsetmacro{\ya}{cos(\phi) * \yarot * (\zo / \za)}
      \pgfmathsetmacro{\xa}{\xarot * (\zo / \za)}    

      \pgfmathsetmacro{\xbnom}{\xo + \Lx}
      \pgfmathsetmacro{\ybnom}{\yo + (\n / 5) * \Ly)}
    
      \pgfmathsetmacro{\xbrot}{cos(\theta) * \xbnom + sin(\theta) * \ybnom}
      \pgfmathsetmacro{\ybrot}{-sin(\theta) * \xbnom + cos(\theta) * \ybnom}
    
      \pgfmathsetmacro{\zb}{\zo + sin(\phi) * \ybrot}
      \pgfmathsetmacro{\yb}{cos(\phi) * \ybrot * (\zo / \zb)}
      \pgfmathsetmacro{\xb}{\xbrot * (\zo / \zb)}   
    
      \draw[gray70, line width=0.25] ({\xa}, {\ya}) -- ({\xb}, {\yb});
    }
  
    \foreach \n in {-4, -3, ..., 4} {
      \pgfmathsetmacro{\xanom}{\xo + (\n / 5) * \Lx}
      \pgfmathsetmacro{\yanom}{\yo - \Ly}
    
      \pgfmathsetmacro{\xarot}{cos(\theta) * \xanom + sin(\theta) * \yanom}
      \pgfmathsetmacro{\yarot}{-sin(\theta) * \xanom + cos(\theta) * \yanom}
    
      \pgfmathsetmacro{\za}{\zo + sin(\phi) * \yarot}
      \pgfmathsetmacro{\ya}{cos(\phi) * \yarot * (\zo / \za)}
      \pgfmathsetmacro{\xa}{\xarot * (\zo / \za)}      

      \pgfmathsetmacro{\xbnom}{\xo + (\n / 5) * \Lx}
      \pgfmathsetmacro{\ybnom}{\yo + \Ly}
    
      \pgfmathsetmacro{\xbrot}{cos(\theta) * \xbnom + sin(\theta) * \ybnom}
      \pgfmathsetmacro{\ybrot}{-sin(\theta) * \xbnom + cos(\theta) * \ybnom}
    
      \pgfmathsetmacro{\zb}{\zo + sin(\phi) * \ybrot}
      \pgfmathsetmacro{\yb}{cos(\phi) * \ybrot * (\zo / \zb)}
      \pgfmathsetmacro{\xb}{\xbrot * (\zo / \zb)}   
    
      \draw[gray70, line width=0.25] ({\xa}, {\ya}) -- ({\xb}, {\yb});
    }
  \end{scope}
  
  \draw[mid, dashed] (\bxa, \bya) -- (\bxb, \byb) -- (\bxc, \byc) -- (\bxd, \byd) -- cycle;
  \node[color=mid] at (8, 0) { $T_{\psi(q)} Q$ };
  
  \pgfmathsetmacro{\xnom}{2}
  \pgfmathsetmacro{\ynom}{0}
    
  \pgfmathsetmacro{\xrot}{cos(\theta) * \xnom + sin(\theta) * \ynom}
  \pgfmathsetmacro{\yrot}{-sin(\theta) * \xnom + cos(\theta) * \ynom}
    
  \pgfmathsetmacro{\z}{\zo + sin(\phi) * \yrot}
  \pgfmathsetmacro{\y}{cos(\phi) * \yrot * (\zo / \z)}
  \pgfmathsetmacro{\x}{\xrot * (\zo / \z)}  
  
  \draw[color=mid, ->, >=stealth] (0, 0) -- (\x, \y);
  
  \pgfmathsetmacro{\xnom}{0}
  \pgfmathsetmacro{\ynom}{1}
    
  \pgfmathsetmacro{\xrot}{cos(\theta) * \xnom + sin(\theta) * \ynom}
  \pgfmathsetmacro{\yrot}{-sin(\theta) * \xnom + cos(\theta) * \ynom}
    
  \pgfmathsetmacro{\z}{\zo + sin(\phi) * \yrot}
  \pgfmathsetmacro{\y}{cos(\phi) * \yrot * (\zo / \z)}
  \pgfmathsetmacro{\x}{\xrot * (\zo / \z)}  
  
  \draw[color=mid, ->, >=stealth] (0, 0) -- (\x, \y);
  
  \fill [fill=dark] (0, 0) circle (0.2);
  \node[color=dark] at (0.5, -1) { $\psi(q)$ };

\end{tikzpicture}
}
\caption{
A reparameterization $\psi: Q \rightarrow Q$ transforms not only points in 
the manifold but also geometric objects defined in the tangent and cotangent
spaces.  Locally the action of a reparameterization behaves like a rotation
given by the Jacobian matrix of the reparameterizing map.
}
\label{fig:complete_reparameterization} 
\end{figure*}

Importantly a reparameterization of the base manifold also affects the local structure 
of the tangent and cotangent spaces (Figure \ref{fig:complete_reparameterization}).  
Tangent vectors push forward along the transformation, the vector $v \in T_{q} Q$ mapping 
into a vector $v' \in T_{\psi(q)} Q$.  Local bases of a tangent spaces transform as 
\begin{equation*}
(\partial')_{i} = J^{j}_{i} (\psi^{-1}(q')) \cdot \partial_{j},
\end{equation*}
which immediately implies that the components of a vector in that basis transforms to the 
components 
\begin{equation*}
(v')^{i} = J^{i}_{j} (\psi^{-1}(q')) \cdot v^{j}.
\end{equation*}
Likewise probability density functions over a tangent space acquire the same Jacobian 
determinant as the probability density functions in the local charts.  This implies that 
conditional probability density functions over the tangent bundle pick up \emph{two} 
factors of the inverse Jacobian determinant,
\begin{equation*}
\pi(q', v') = \pi(\psi^{-1}(q'), J^{i}_{j}(\psi^{-1}(q')) \, (v')^{j}) 
\cdot \left| J (\psi^{-1}(q')) \right|^{-1} \cdot \left| J (\psi^{-1}(q')) \right|^{-1}.
\end{equation*}

At the same time we can take a more comprehensive perspective and note that any 
reparameterization over the base manifold $Q$ induces a reparameterization of 
the entire tangent bundle at once.  The Jacobian matrix of this bundle 
reparameterization is block diagonal with both blocks equal to the Jacobian
matrix of the base reparameterization, 
\begin{equation*}
J_{TQ} = 
\begin{pmatrix}
J & 0 \\
0 & J
\end{pmatrix},
\end{equation*}
from which one can readily reproduce all of the previous results.  For 
example,
\begin{equation*}
\left| J_{TQ} \right| = \left| \begin{pmatrix}
J & 0 \\
0 & J
\end{pmatrix} \right| = J^{2}.
\end{equation*}

Objects in the cotangent spaces naturally pull back along the reparameterization 
and hence transform in the opposite way as tangent vectors.  A local basis of a 
cotangent space transforms as
\begin{equation*}
(\mathrm{d}q')^{i} = J^{i}_{j} (\psi^{-1}(q')) \cdot \mathrm{d}q^{j},
\end{equation*}
and the component of a covector in that basis transform as 
\begin{equation*}
(p')_{i} = (J^{-1})^{j}_{i} (\psi^{-1}(q')) \cdot p_{j}.
\end{equation*}
Probability density functions over a cotangent space behave opposite to probability 
density functions over a local chart or in a tangent space; they acquire a factor 
of the Jacobian determinant without inversion.  Critically this implies that probability 
density functions over the cotangent bundle pick up \emph{no} Jacobian factors under a 
reparameterization
\begin{align*}
\pi(q', p') 
&= 
\pi(\psi^{-1}(q'), (J^{-1})^{j}_{i} (\psi^{-1}(q')) \, (p')_{j}) 
\cdot \left| J (\psi^{-1}(q')) \right|^{-1} \cdot \left| J (\psi^{-1}(q')) \right|
\\
&=
\pi(\psi^{-1}(q'), (J^{-1})^{j}_{i} (q) \, (p')_{j}).
\end{align*}
This hints at the natural probabilistic structure of the cotangent bundle and some of
inherent advantages of algorithms like Langevin Monte Carlo and Hamiltonian Monte Carlo
defined there.

Just as a reparameterization of the base manifold induces a reparameterization of the 
tangent bundle, it also induces a reparameterization of the entire cotangent bundle.  
Here the Jacobian matrix of this bundle reparameterization is block diagonal,
but the lower block now equals the inverse Jacobian matrix of the base reparameterization,
*
\begin{equation*}
J_{T^{*}Q} = 
\begin{pmatrix}
J & 0 \\
0 & J^{-1}
\end{pmatrix}.
\end{equation*}
This perspective makes it particularly clear that the Jacobian determinant of the induced 
reparameterization is exactly one,
\begin{equation*}
\left| J_{T^{*}Q} \right| = \left| \begin{pmatrix}
J & 0 \\
0 & J^{-1}
\end{pmatrix} \right| = J \cdot J^{-1} = 1.
\end{equation*}

From the transformation properties of vectors and covectors we can work out how general 
tensors transform.  In particular we can work out how the components functions of a 
Riemannian metric transform when we reparameterize the base space.  In this case we get 
two inverse Jacobians, one for each component,
\begin{equation*}
(g')_{lm}(q') = 
(J^{-1})^{i}_{l} (\psi^{-1}(q'))
\cdot
(J^{-1})^{j}_{m} (\psi^{-1}(q'))
\cdot g_{ij}(\psi^{-1}(q')).
\end{equation*}
As we'd expect from a geometric invariant, the quadratic form defining an elliptical 
probability density function on the tangent spaces doesn't change under the reparameterization,
\begin{align*}
(g')_{q'}(v', v')
&= 
(g')_{lm}(q') (v')^{l} (v')^{m}
\\
&= \;
(J^{-1})^{i}_{l} (\psi^{-1}(q'))
\cdot
(J^{-1})^{j}_{m} (\psi^{-1}(q'))
\cdot g_{ij}(q)
\\
& \quad \cdot
J^{l}_{r} (\psi^{-1}(q')) \cdot v^{r}
\\
& \quad \cdot
J^{m}_{s} (\psi^{-1}(q')) \cdot v^{s}
\\
&= \;
\Big[
(J^{-1})^{i}_{l} (\psi^{-1}(q'))
\cdot
J^{l}_{r} (\psi^{-1}(q')) \Big] 
\cdot
\Big[
J^{j}_{m} (\psi^{-1}(q'))
\cdot
(J^{-1})^{m}_{s} (\psi^{-1}(q')) \Big]
\\
& \quad \cdot g_{ij}(q) \cdot v^{r} \cdot v^{s}
\\
&=
\Big[ \delta^{i}_{r} \Big] \cdot \Big[ \delta^{j}_{s} \Big]
\cdot g_{ij}(q) \cdot v^{r} \cdot v^{s}
\\
&=
g_{ij}(q) \cdot v^{i} \cdot v^{j}
\\
&=
g_{q}(v, v).
\end{align*}
The transformation properties of the elliptical probability density functions instead 
depend entirely on their metric determinant terms.

Provided that we reparameterize not just the base space but also the tangent and cotangent 
spaces, then any algorithm based on exact flows will be invariant to reparameterizations; 
the entire Markov chains they generate will map forward from one parameterization to another 
without any changes to their dynamics.  Algorithms that depend on discrete approximations
to these flows will not be exactly invariant, as the approximation error and any correction 
scheme will in general depend on the local parameterization, but the resulting dynamics will 
be similar.

Reparameterization, however, is often recommended to improve performance because it is 
\emph{supposed} to change the dynamics of the Markov chain.  This contradiction is resolved 
when we realize that the reparameterizations employed in practice are not the complete 
reparameterizations of a proper geometric system but rather incomplete reparameterizations 
that transform the initial system into something else entirely.

\subsection{Incomplete Reparameterizations and Equivalent Metrics} \label{sec:equivalent_metric}

In practice any geometric algorithm is implemented with coordinates, components, and 
probability densities.  Typically, however, only the target probability density is 
exposed to the user.  Conditional probability density functions on the tangent or 
cotangent bundle, or components of the metric that define those conditional densities, 
are set to default values not exposed to the user or exposed but significantly limited 
in flexibility.  For example, the default configuration of Stan \citep{Stan:2019}
forces a metric with constant, diagonal components.

Consequently a user cannot reparameterize the entire geometric system on which these 
algorithms are based.  They can only reparameterize the target probability density 
function while the tangent and cotangent structures remain \emph{fixed}.  These 
incomplete reparameterizations result in a \emph{different} geometry and hence a 
different algorithm that may interact better or worse with the target distribution.

\begin{figure*}
\centering
\subfigure[]{
\begin{tikzpicture}[scale=0.35, thick,
declare function={ fr(\t) = 3 * (1 - \t) * (1 - \t) * \t * 4
                           + 3 * (1 - \t) * \t * \t * 5 + \t * \t * \t * 8;},
declare function={ fh(\t) =  3 * (1 - \t) * \t * \t * 6 + \t * \t * \t * 8;} 
]
  \draw[color=white] (-10.25, 0) -- (10.25, 0);

  \draw[dashed] (0, -8) circle [x radius=8, y radius={8 * sin(7)}]; 

  \filldraw[draw=black, fill=white] (-7.92, -7.85) .. controls (-5, -6) and (-4, 0.1) 
    .. (0, 0.1) .. controls (4, 0.1) and (5, -6) .. (7.92, -7.85);
    
  \def\radii{{8, 6.85, 6.05, 5.435, 4.845, 4.175, 3.45, 2.5}}
  \def\heights{{8, 7, 6, 5, 4, 3, 2, 1}}
  \foreach \n in {0, 1, ..., 7} {
    \pgfmathparse{\radii[\n]}\edef\r{\pgfmathresult}
    \pgfmathparse{\heights[\n]}\edef\h{\pgfmathresult}
    \draw[line width=0.5, color=black, ->, >=stealth] ({0 - \r * cos(7) * sin(15)}, -\h) arc(255:210:{\r} and {\r * sin(7)}); 
  }
  
  \node[] at (9, -7) { $Q$ };
  
  \pgfmathsetmacro{\xo}{0}
  \pgfmathsetmacro{\yo}{0}
  \pgfmathsetmacro{\zo}{5}

  \pgfmathsetmacro{\bLx}{5}
  \pgfmathsetmacro{\bLy}{2.5}
  
  \pgfmathsetmacro{\Lx}{5}
  \pgfmathsetmacro{\Ly}{5}
  
  \pgfmathsetmacro{\dxo}{5}
  \pgfmathsetmacro{\dyo}{0.5}
  
  \pgfmathsetmacro{\theta}{0}
  \pgfmathsetmacro{\phi}{60}

  \pgfmathsetmacro{\xnom}{\xo - \bLx}
  \pgfmathsetmacro{\ynom}{\yo - \bLy}
    
  \pgfmathsetmacro{\z}{\zo + sin(\phi) * \ynom}
  \pgfmathsetmacro{\bya}{cos(\phi) * \ynom * (\zo / \z)}
  \pgfmathsetmacro{\bxa}{\xnom * (\zo / \z)}    

  \pgfmathsetmacro{\xnom}{\xo - \bLx}
  \pgfmathsetmacro{\ynom}{\yo + \bLy}
    
  \pgfmathsetmacro{\z}{\zo + sin(\phi) * \ynom}
  \pgfmathsetmacro{\byb}{cos(\phi) * \ynom * (\zo / \z)}
  \pgfmathsetmacro{\bxb}{\xnom * (\zo / \z)}  
  
  \pgfmathsetmacro{\xnom}{\xo + \bLx}
  \pgfmathsetmacro{\ynom}{\yo + \bLy}
    
  \pgfmathsetmacro{\z}{\zo + sin(\phi) * \ynom}
  \pgfmathsetmacro{\byc}{cos(\phi) * \ynom * (\zo / \z)}
  \pgfmathsetmacro{\bxc}{\xnom * (\zo / \z)}      

  \pgfmathsetmacro{\xnom}{\xo + \bLx}
  \pgfmathsetmacro{\ynom}{\yo - \bLy}

  \pgfmathsetmacro{\z}{\zo + sin(\phi) * \ynom}
  \pgfmathsetmacro{\byd}{cos(\phi) * \ynom * (\zo / \z)}
  \pgfmathsetmacro{\bxd}{\xnom * (\zo / \z)} 
  
  \fill[mid, color=mid, opacity=0.10] (\bxa, \bya) -- (\bxb, \byb) -- (\bxc, \byc) -- (\bxd, \byd) -- cycle;

  \begin{scope} 
    \clip (\bxa, \bya) -- (\bxb, \byb) -- (\bxc, \byc) -- (\bxd, \byd) -- cycle;

    \foreach \n in {-4, -3, ..., 4} {
      \pgfmathsetmacro{\xanom}{\xo - \Lx}
      \pgfmathsetmacro{\yanom}{\yo + (\n / 5) * \Ly)}
    
      \pgfmathsetmacro{\xarot}{cos(\theta) * \xanom + sin(\theta) * \yanom}
      \pgfmathsetmacro{\yarot}{-sin(\theta) * \xanom + cos(\theta) * \yanom}
    
      \pgfmathsetmacro{\za}{\zo + sin(\phi) * \yarot}
      \pgfmathsetmacro{\ya}{cos(\phi) * \yarot * (\zo / \za)}
      \pgfmathsetmacro{\xa}{\xarot * (\zo / \za)}    

      \pgfmathsetmacro{\xbnom}{\xo + \Lx}
      \pgfmathsetmacro{\ybnom}{\yo + (\n / 5) * \Ly)}
    
      \pgfmathsetmacro{\xbrot}{cos(\theta) * \xbnom + sin(\theta) * \ybnom}
      \pgfmathsetmacro{\ybrot}{-sin(\theta) * \xbnom + cos(\theta) * \ybnom}
    
      \pgfmathsetmacro{\zb}{\zo + sin(\phi) * \ybrot}
      \pgfmathsetmacro{\yb}{cos(\phi) * \ybrot * (\zo / \zb)}
      \pgfmathsetmacro{\xb}{\xbrot * (\zo / \zb)}   
    
      \draw[gray70, line width=0.25] ({\xa}, {\ya}) -- ({\xb}, {\yb});
    }
  
    \foreach \n in {-4, -3, ..., 4} {
      \pgfmathsetmacro{\xanom}{\xo + (\n / 5) * \Lx}
      \pgfmathsetmacro{\yanom}{\yo - \Ly}
    
      \pgfmathsetmacro{\xarot}{cos(\theta) * \xanom + sin(\theta) * \yanom}
      \pgfmathsetmacro{\yarot}{-sin(\theta) * \xanom + cos(\theta) * \yanom}
    
      \pgfmathsetmacro{\za}{\zo + sin(\phi) * \yarot}
      \pgfmathsetmacro{\ya}{cos(\phi) * \yarot * (\zo / \za)}
      \pgfmathsetmacro{\xa}{\xarot * (\zo / \za)}      

      \pgfmathsetmacro{\xbnom}{\xo + (\n / 5) * \Lx}
      \pgfmathsetmacro{\ybnom}{\yo + \Ly}
    
      \pgfmathsetmacro{\xbrot}{cos(\theta) * \xbnom + sin(\theta) * \ybnom}
      \pgfmathsetmacro{\ybrot}{-sin(\theta) * \xbnom + cos(\theta) * \ybnom}
    
      \pgfmathsetmacro{\zb}{\zo + sin(\phi) * \ybrot}
      \pgfmathsetmacro{\yb}{cos(\phi) * \ybrot * (\zo / \zb)}
      \pgfmathsetmacro{\xb}{\xbrot * (\zo / \zb)}   
    
      \draw[gray70, line width=0.25] ({\xa}, {\ya}) -- ({\xb}, {\yb});
    }
  \end{scope}
  
  \draw[mid, dashed] (\bxa, \bya) -- (\bxb, \byb) -- (\bxc, \byc) -- (\bxd, \byd) -- cycle;
  \node[color=mid] at (8, 0) { $T_{\psi(q)} Q$ };
   
  \pgfmathsetmacro{\xnom}{2}
  \pgfmathsetmacro{\ynom}{0}
    
  \pgfmathsetmacro{\xrot}{cos(\theta) * \xnom + sin(\theta) * \ynom}
  \pgfmathsetmacro{\yrot}{-sin(\theta) * \xnom + cos(\theta) * \ynom}
    
  \pgfmathsetmacro{\z}{\zo + sin(\phi) * \yrot}
  \pgfmathsetmacro{\y}{cos(\phi) * \yrot * (\zo / \z)}
  \pgfmathsetmacro{\x}{\xrot * (\zo / \z)}  
  
  \draw[color=mid, ->, >=stealth] (0, 0) -- (\x, \y);
  
  \pgfmathsetmacro{\xnom}{0}
  \pgfmathsetmacro{\ynom}{1}
    
  \pgfmathsetmacro{\xrot}{cos(\theta) * \xnom + sin(\theta) * \ynom}
  \pgfmathsetmacro{\yrot}{-sin(\theta) * \xnom + cos(\theta) * \ynom}
    
  \pgfmathsetmacro{\z}{\zo + sin(\phi) * \yrot}
  \pgfmathsetmacro{\y}{cos(\phi) * \yrot * (\zo / \z)}
  \pgfmathsetmacro{\x}{\xrot * (\zo / \z)}  
  
  \draw[color=mid, ->, >=stealth] (0, 0) -- (\x, \y);
  
  \fill [fill=dark] (0, 0) circle (0.2);
  \node[color=dark] at (0.5, -1) { $\psi(q)$ }; 

\end{tikzpicture}
}
\subfigure[]{
\begin{tikzpicture}[scale=0.35, thick,
declare function={ fr(\t) = 3 * (1 - \t) * (1 - \t) * \t * 4
                           + 3 * (1 - \t) * \t * \t * 5 + \t * \t * \t * 8;},
declare function={ fh(\t) =  3 * (1 - \t) * \t * \t * 6 + \t * \t * \t * 8;} 
]
  \draw[color=white] (-10.25, 0) -- (10.25, 0); 

  \draw[dashed] (0, -8) circle [x radius=8, y radius={8 * sin(7)}]; 

  \filldraw[draw=black, fill=white] (-7.92, -7.85) .. controls (-5, -6) and (-4, 0.1) 
    .. (0, 0.1) .. controls (4, 0.1) and (5, -6) .. (7.92, -7.85);
    
  \def\radii{{8, 6.85, 6.05, 5.435, 4.845, 4.175, 3.45, 2.5}}
  \def\heights{{8, 7, 6, 5, 4, 3, 2, 1}}
  \foreach \n in {0, 1, ..., 7} {
    \pgfmathparse{\radii[\n]}\edef\r{\pgfmathresult}
    \pgfmathparse{\heights[\n]}\edef\h{\pgfmathresult}
    \draw[line width=0.5, color=black, <-, >=stealth] ({0 - \r * cos(7) * sin(15)}, -\h) arc(255:210:{\r} and {\r * sin(7)}); 
  }
  
  \node[] at (9, -7) { $Q$ };
  
  \pgfmathsetmacro{\xo}{0}
  \pgfmathsetmacro{\yo}{0}
  \pgfmathsetmacro{\zo}{5}

  \pgfmathsetmacro{\bLx}{5}
  \pgfmathsetmacro{\bLy}{2.5}
  
  \pgfmathsetmacro{\Lx}{5}
  \pgfmathsetmacro{\Ly}{5}
  
  \pgfmathsetmacro{\dxo}{5}
  \pgfmathsetmacro{\dyo}{0.5}
  
  \pgfmathsetmacro{\theta}{-30}
  \pgfmathsetmacro{\phi}{60}

  \pgfmathsetmacro{\xnom}{\xo - \bLx}
  \pgfmathsetmacro{\ynom}{\yo - \bLy}
    
  \pgfmathsetmacro{\z}{\zo + sin(\phi) * \ynom}
  \pgfmathsetmacro{\bya}{cos(\phi) * \ynom * (\zo / \z)}
  \pgfmathsetmacro{\bxa}{\xnom * (\zo / \z)}    

  \pgfmathsetmacro{\xnom}{\xo - \bLx}
  \pgfmathsetmacro{\ynom}{\yo + \bLy}
    
  \pgfmathsetmacro{\z}{\zo + sin(\phi) * \ynom}
  \pgfmathsetmacro{\byb}{cos(\phi) * \ynom * (\zo / \z)}
  \pgfmathsetmacro{\bxb}{\xnom * (\zo / \z)}  
  
  \pgfmathsetmacro{\xnom}{\xo + \bLx}
  \pgfmathsetmacro{\ynom}{\yo + \bLy}
    
  \pgfmathsetmacro{\z}{\zo + sin(\phi) * \ynom}
  \pgfmathsetmacro{\byc}{cos(\phi) * \ynom * (\zo / \z)}
  \pgfmathsetmacro{\bxc}{\xnom * (\zo / \z)}      

  \pgfmathsetmacro{\xnom}{\xo + \bLx}
  \pgfmathsetmacro{\ynom}{\yo - \bLy}

  \pgfmathsetmacro{\z}{\zo + sin(\phi) * \ynom}
  \pgfmathsetmacro{\byd}{cos(\phi) * \ynom * (\zo / \z)}
  \pgfmathsetmacro{\bxd}{\xnom * (\zo / \z)} 
  
  \fill[mid, color=mid, opacity=0.10] (\bxa, \bya) -- (\bxb, \byb) -- (\bxc, \byc) -- (\bxd, \byd) -- cycle;

  \begin{scope} 
    \clip (\bxa, \bya) -- (\bxb, \byb) -- (\bxc, \byc) -- (\bxd, \byd) -- cycle;

    \foreach \n in {-4, -3, ..., 4} {
      \pgfmathsetmacro{\xanom}{\xo - \Lx}
      \pgfmathsetmacro{\yanom}{\yo + (\n / 5) * \Ly)}
    
      \pgfmathsetmacro{\xarot}{cos(\theta) * \xanom + sin(\theta) * \yanom}
      \pgfmathsetmacro{\yarot}{-sin(\theta) * \xanom + cos(\theta) * \yanom}
    
      \pgfmathsetmacro{\za}{\zo + sin(\phi) * \yarot}
      \pgfmathsetmacro{\ya}{cos(\phi) * \yarot * (\zo / \za)}
      \pgfmathsetmacro{\xa}{\xarot * (\zo / \za)}    

      \pgfmathsetmacro{\xbnom}{\xo + \Lx}
      \pgfmathsetmacro{\ybnom}{\yo + (\n / 5) * \Ly)}
    
      \pgfmathsetmacro{\xbrot}{cos(\theta) * \xbnom + sin(\theta) * \ybnom}
      \pgfmathsetmacro{\ybrot}{-sin(\theta) * \xbnom + cos(\theta) * \ybnom}
    
      \pgfmathsetmacro{\zb}{\zo + sin(\phi) * \ybrot}
      \pgfmathsetmacro{\yb}{cos(\phi) * \ybrot * (\zo / \zb)}
      \pgfmathsetmacro{\xb}{\xbrot * (\zo / \zb)}   
    
      \draw[gray70, line width=0.25] ({\xa}, {\ya}) -- ({\xb}, {\yb});
    }
  
    \foreach \n in {-4, -3, ..., 4} {
      \pgfmathsetmacro{\xanom}{\xo + (\n / 5) * \Lx}
      \pgfmathsetmacro{\yanom}{\yo - \Ly}
    
      \pgfmathsetmacro{\xarot}{cos(\theta) * \xanom + sin(\theta) * \yanom}
      \pgfmathsetmacro{\yarot}{-sin(\theta) * \xanom + cos(\theta) * \yanom}
    
      \pgfmathsetmacro{\za}{\zo + sin(\phi) * \yarot}
      \pgfmathsetmacro{\ya}{cos(\phi) * \yarot * (\zo / \za)}
      \pgfmathsetmacro{\xa}{\xarot * (\zo / \za)}      

      \pgfmathsetmacro{\xbnom}{\xo + (\n / 5) * \Lx}
      \pgfmathsetmacro{\ybnom}{\yo + \Ly}
    
      \pgfmathsetmacro{\xbrot}{cos(\theta) * \xbnom + sin(\theta) * \ybnom}
      \pgfmathsetmacro{\ybrot}{-sin(\theta) * \xbnom + cos(\theta) * \ybnom}
    
      \pgfmathsetmacro{\zb}{\zo + sin(\phi) * \ybrot}
      \pgfmathsetmacro{\yb}{cos(\phi) * \ybrot * (\zo / \zb)}
      \pgfmathsetmacro{\xb}{\xbrot * (\zo / \zb)}   
    
      \draw[gray70, line width=0.25] ({\xa}, {\ya}) -- ({\xb}, {\yb});
    }
  \end{scope}
  
  \draw[mid, dashed] (\bxa, \bya) -- (\bxb, \byb) -- (\bxc, \byc) -- (\bxd, \byd) -- cycle;
  \node[color=mid] at (8, 0) { $T_{q} Q$ };
   
  \pgfmathsetmacro{\xnom}{2}
  \pgfmathsetmacro{\ynom}{0}
    
  \pgfmathsetmacro{\xrot}{cos(\theta) * \xnom + sin(\theta) * \ynom}
  \pgfmathsetmacro{\yrot}{-sin(\theta) * \xnom + cos(\theta) * \ynom}
    
  \pgfmathsetmacro{\z}{\zo + sin(\phi) * \yrot}
  \pgfmathsetmacro{\y}{cos(\phi) * \yrot * (\zo / \z)}
  \pgfmathsetmacro{\x}{\xrot * (\zo / \z)}  
  
  \draw[color=mid, ->, >=stealth] (0, 0) -- (\x, \y);
  
  \pgfmathsetmacro{\xnom}{0}
  \pgfmathsetmacro{\ynom}{1}
    
  \pgfmathsetmacro{\xrot}{cos(\theta) * \xnom + sin(\theta) * \ynom}
  \pgfmathsetmacro{\yrot}{-sin(\theta) * \xnom + cos(\theta) * \ynom}
    
  \pgfmathsetmacro{\z}{\zo + sin(\phi) * \yrot}
  \pgfmathsetmacro{\y}{cos(\phi) * \yrot * (\zo / \z)}
  \pgfmathsetmacro{\x}{\xrot * (\zo / \z)}  
  
  \draw[color=mid, ->, >=stealth] (0, 0) -- (\x, \y);
  
  \fill [fill=dark] (0, 0) circle (0.2);
  \node[color=dark] at (0.5, -1) { $q$ }; 

\end{tikzpicture}
}
\caption{
An incomplete reparameterization forces the metric geometry to be
defined using component functions in the reparameterized charts, 
not the initial charts.  Transforming back to the initial 
parameterization we see that this is equivalent to defining a 
different metric on the original space that we might have anticipated.  
In other words an incomplete reparameterization transforms the base 
manifold while holding the tangent and cotangent spaces fixed, 
twisting the tangent and cotangent bundles.  If we release our 
hold on these spaces the bundles snap back, revealing the equivalent 
metric geometries.
}
\label{fig:incomplete_reparameterization} 
\end{figure*}
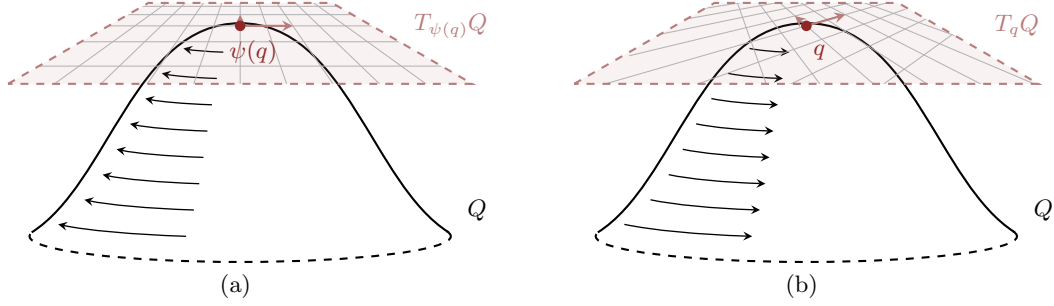

While an incomplete reparameterization is not a proper geometric transformation, its
effect does admit a convenient geometric interpretation.  We begin with a metric 
specified by the local coordinate functions $g_{ij}(q)$.  Reparameterizing the base
manifold, $q \mapsto q' = \phi(q)$, but fixing the components of the metric
results in a \emph{new} metric specified by the same components but in the new
coordinate system, $g_{lm}(q')$.  To compare these metrics we have to completely invert 
the reparameterization, pulling the new metric back into the original coordinate system,
\begin{align*}
g_{ij}(q) 
&= 
(J)^{l}_{i} (\psi^{-1}(q'))
\cdot
(J)^{m}_{j} (\psi^{-1}(q'))
\cdot g_{lm}(\psi^{-1}(q'))
\\
&=
(J)^{l}_{i} (q)
\cdot
(J)^{m}_{j} (q)
\cdot g_{lm}(q).
\end{align*}
In words, an incomplete reparameterization is equivalent to running the algorithm in 
the original coordinate system but with the transformed metric
(Figure \ref{fig:incomplete_reparameterization})
\begin{equation*}
\bar{g}_{ij}(q) 
=
(J)^{l}_{i} (q)
\cdot
(J)^{m}_{j} (q)
\cdot g_{lm}(q).
\end{equation*}
Consequently there is a one-to-one equivalence between incomplete reparameterizations 
and the choice of metric, and hence the configuration of a Riemannian Markov transition.

Under an incomplete reparameterization geodesics and elliptical conditional 
probability distributions on the tangent and cotangent spaces all follow 
from this new, equivalent metric.  These new configurations will induce new 
dynamics with respect to the target distribution, resulting in modified 
performance that may or may not be beneficial.

One advantage of this equivalence is that by applying an incomplete 
reparameterization we can effectively implement an algorithm with 
spatially-varying metric components using only an algorithm configuration
with constant metric components, at least if we can find the right 
reparameterization.  Riemannian Markov transitions in a coordinate system 
admitting constant metric components are significantly easier to robustly 
implement and indeed are often the only option in popular software packages. 
For example we can use explicit symplectic integrators in to implement 
Langevin and Hamiltonian Monte Carlo instead of more expensive, and more
fragile, implicit symplectic integrators.  By finding an appropriate 
reparameterization we can reproduce the geometry of a more sophisticated 
metric without modifying the software itself. 

Algorithms exploiting metrics with spatially-varying components in the 
default coordinate system are often denoted ``Riemannian'' algorithms in 
the statistics literature, with those using constant components denoted 
``Euclidean'' algorithms.  Using this terminology an incomplete
reparameterization allows one to effectively run a Riemannian algorithm 
using only a Euclidean implementation. Keep in mind, however, that this 
terminology is technically incorrect, as discussed at the end of Section 
\ref{sec:metrics}.

\subsection{Optimal Incomplete Reparameterizations}

An immediate advantage of this identification between an incomplete 
reparameterization and its equivalent Riemannian geometry is that it allows 
us to determine reparameterizations that optimize performance with respect 
to a given target distribution by first determining the optimal Riemannian 
geometry.

Within a small neighborhood we can approximate log target probability density
function with a Taylor expansion, although that approximation has to be made 
with care.  Firstly the target probability density function is not an invariant 
function amenable to approximation.  We can construct an appropriate function,
however, by using the determinant of the metric to correct for the non-invariant 
behaviors,
\begin{align*}
\rho(q) 
&= \log \left( \pi(q) \cdot |g(q)|^{-\frac{1}{2} } \right)
\\
&= \log \pi(q) - \frac{1}{2} \log |g(q)|.
\end{align*}
We can then we can construct a local Taylor expansion of this invariant function,
\begin{equation*}
\rho(q) = 
\rho(q_{0}) + 
\frac{\partial \rho}{\partial q^{i} }(q_{0}) \cdot (q - q_{0})^{i}
+ \frac{1}{2} 
\frac{\partial^{2} \rho}{\partial q^{i} \partial q^{j} }(q_{0}) 
\cdot (q - q_{0})^{i} \cdot (q - q_{0})^{j} + \ldots.
\end{equation*}
Finally if the local chart is in a basin where the gradient, as well as all of 
the higher-order terms, are negligible compared to the constant and quadratic
terms then we can approximate the function as
\begin{equation*}
\rho(q) \approx 
\mathrm{const}
+ \frac{1}{2} 
\frac{\partial^{2} \rho}{\partial q^{i} \partial q^{j} }(q_{0}) 
\cdot (q - q_{0})^{i} \cdot (q - q_{0})^{j} 
\end{equation*}
If the second derivatives of this function in that small neighborhood are all 
positive then this approximation defines a Gaussian probability density function.
In other words in sufficiently small charts where $\rho(q)$ is concave we 
can approximate the target probability density function with a multivariate Gaussian 
probability density function defined by the precision matrix
\begin{equation*}
(\Sigma^{-1})_{ij} =
\frac{\partial^{2} \rho}{\partial q^{i} \partial q^{j} }(q_{0}).
\end{equation*}

This approximation significantly simplifies the analysis of geometric algorithms 
that utilize elliptical probability density functions over the tangent or cotangent 
spaces.  Rotating the entire tangent bundle within the small neighborhood, for 
example, exchanges covariance between the target approximation and the covariance 
of the tangent probability density function defined by the metric.  At the same time 
rotating the cotangent bundle exchanges precision between the target approximation
and the cotangent probability density function defined by the inverse metric.
Consequently we can completely decorrelate the local approximation to the target
probability density function by choosing a metric that compensates for the local 
behavior of the Hessian of $\rho$.  In each neighborhood this would be accomplished 
with a metric specified by the components
\begin{equation*}
g_{ij}(q_{0}) 
= 
\frac{\partial^{2} \rho}{\partial q^{i} \partial q^{j} }(q_{0}).
\end{equation*}

Unfortunately this equality is valid only within a single chart and hence does not 
define an optimization criterion that is consistent across the entire base manifold.  
The main problem is that the Hessian does not transform like a metric but rather a 
\emph{jet}, in particular a one-dimensional, rank-two covelocity \citep{Betancourt:2018b}.  
We can use the Riemannian structure on the manifold, however, to correct the Hessian 
into a geometric object that we can compare to the metric.

The \emph{covariant Hessian} uses the linear connection to compensate for the 
non-tensorial behavior of the Hessian,
\begin{equation*}
\nabla^{2} f(q) 
= 
\left( \frac{ \partial^{2} f }{ \partial q^{i} \partial q^{j} } (q)
+ \Gamma^{k}_{ij}(q) \frac{ \partial f}{ \partial q^{k} }(q) \right) 
\mathrm{d} q^{i} \otimes \mathrm{d} q^{j},
\end{equation*}
consistently across all charts.  Local comparisons between the covariant Hessian and 
the metric are then self-consistent across the entire base manifold.  This allows us
to construct a proper criterion for metric optimality at each point as
\begin{align*}
g_{ij}(q_{0}) 
&=
\nabla^{2}_{ij} \lambda(q_{0})
\\
&=
\nabla^{2}_{ij} (\log \pi - \frac{1}{2} \log |g| )(q_{0}),
\end{align*}
or, using the fact that the covariant Hessian of any
function of the metric vanishes,
\begin{align*}
g_{ij}(q_{0}) 
&=
\nabla^{2}_{ij} (\log \pi - \frac{1}{2} \log |g| )(q_{0})
\\
&=
\nabla^{2}_{ij} \log \pi(q_{0})
\\
&=
\frac{ \partial^{2} \log \pi }{ \partial q^{i} \partial q^{j} } (q_{0})
+ \Gamma^{k}_{ij} \frac{ \partial \log \pi}{ \partial q^{k} } (q_{0}).
\end{align*}

Likewise the local deviation from optimality can be quantified by the 
difference
\begin{align*}
\Delta(q) &= g(q) - \nabla^{2} \log \pi(q).
\end{align*}
We can summarize this deviation with any matrix scalar, for example the 
scalar determinant, $\left| \Delta (q) \right|$.

In practice we can achieve $\Delta(q) = 0$ with the proper choice of metric 
components, but we can also achieve it with an appropriate incomplete 
reparameterization and its equivalent metric,
\begin{equation*}
\bar{g}_{ij}(q) = (J)^{l}_{i} (q)
\cdot
(J)^{m}_{j} (q)
\cdot g_{lm}(q).
\end{equation*}
Consequently substituting the equivalent metric, $\bar{g}_{ij}(q)$ into the 
geometric optimality criterion immediately defines an optimality criterion for 
reparameterizations,
\begin{equation*}
\bar{\Delta}(q) = \bar{g}(q) - \bar{\nabla}^{2} \log \pi(q).
\end{equation*}

Initial excitement is quickly tempered once we inspect the criterion a bit more carefully.  
The criterion defines a system of coupled, non-ordinary differential equations for the 
elements of the Jacobian matrix which define the optimal reparameterization.  This then 
sets up a system of partial differential equations for the optimal reparameterization itself.  
In other words, we will not be solving for optimal reparameterizations in general systems 
any time soon!

The criterion does, however, allow us to analyze specific reparameterizations.  Given a 
specific reparameterization we can verify optimality by constructing the equivalent metric, 
its corresponding connection, and then computing the scalar deviation function, 
$\left| \Delta (q) \right|$.  If the determinant doesn't vanish then we can analyze the 
components of the deviation tensor for insight about the limitations of the chosen 
reparameterization and potential improvements.  Ultimately this criterion provides
the theoretical foundation upon which we can begin formal studies of reparameterizations 
in earnest.

\section{Optimal Reparameterization of Latent Gaussian Models}

To demonstrate the utility of the geometric analysis of incomplete reparameterizations
let's consider the popular reparameterization that arises when transforming from the 
centered parameterization to the non-centered parameterization of a latent Gaussian 
model \citep{PapaspiliopoulosEtAl:2007}.  This reparameterization is known to drastically 
improve the empirical performance of geometric algorithms \citep{BetancourtEtAl:2015} and
we can use our new geometric analysis to provide a more formal motivation for its benefits.

A latent Gaussian model captures the behavior of an unobserved exchangeable population of
individual parameters,
\begin{equation*}
\boldsymbol{\theta} = \left\{ \theta_{1}, \ldots, \theta_{N} \right\},
\end{equation*}
that follow a Gaussian distribution with location $\mu$ and scale $\tau$.  There are 
two natural parameterizations of the individual parameters, and hence two natural 
parameterizations of the latent Gaussian model.  Both parameterizations span the
entire manifold, so we can limit our consideration to the entire space instead of
a single local chart.

\subsection{The Centered Parameterization}

The parameters $\boldsymbol{\theta}$ and $\left\{ \mu, \tau \right\}$ define the 
\emph{centered} parameterization of a latent Gaussian model where the model is specified 
by the probability density function
\begin{align*}
\pi(\boldsymbol{\theta}, \mu, \tau)
&=
\pi(\boldsymbol{\theta} \mid \mu, \tau)
\cdot \pi(\mu, \tau)
\\
&=
\prod_{n = 1}^{N} \mathcal{N} (\theta_{n} \mid \mu, \tau)
\cdot \pi(\mu, \tau).
\end{align*}

Complementing the latent Gaussian model with an observational model for data generated 
from each individual $\theta_{n}$ yields the joint model
\begin{align*}
\pi(\boldsymbol{y}, \boldsymbol{\theta}, \mu, \tau, \phi)
&=
\prod_{n = 1}^{N} \pi (y_{n} \mid \theta_{n}, \phi) \cdot
\prod_{n = 1}^{N} \mathcal{N} (\theta_{n} \mid \mu, \tau)
\cdot \pi(\mu, \tau)
\\
&=
\prod_{n = 1}^{N} \pi (y_{n} \mid \theta_{n}, \phi) \, 
\mathcal{N} (\theta_{n} \mid \mu, \tau)
\cdot \pi(\mu, \tau).
\end{align*}

\begin{figure*}
\centering
\includegraphics[width=2in]{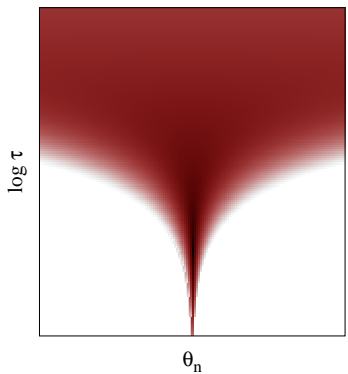}
\caption{
In a centered parameterization the probability density function for a latent
Gaussian model manifests a funnel geometry, where the density concentrates 
into a narrow volume around $\mu$ for small $\tau$ but disperses for large 
$\tau$.  Here and in subsequent figures $\mu$ is fixed at zero.  In order to 
quantify the entire probability distribution a Markov transition must be able 
to explore both regions reasonably quickly which is much easier said than done.
}
\label{fig:funnel} 
\end{figure*}

If the individual likelihood functions are only weakly informative then this joint 
model is dominated by the latent Gaussian probability density function which frustrates 
accurate computation.  The problem is that the interaction between the individual 
parameters and the population scale manifests with a \emph{funnel} geometry.  For 
large $\tau$ the individual $\theta_{n}$ are only weakly coupled to the population 
mean, but for small $\tau$ the $\theta_{n}$ collapse into a narrow concentration 
around $\mu$ (Figure \ref{fig:funnel}).  This rapidly varying curvature frustrates
Markov transitions that cannot dynamically adapt.

On the other hand, as the observational model becomes more informative the individual
likelihood functions concentrate around just those model configurations that are 
consistent the observed data.  Eventually this suppresses the pathological neck of 
the funnel geometry.  Consequently with enough data the posterior probability density 
function will have little contribution from the pathological geometry of the latent 
Gaussian model, and it will be much easier to fit with most Markov transitions.

\subsection{The Non-Centered Parameterization}

The \emph{non-centered} parameterization takes advantage of the fact that we can 
decouple any Gaussian probability density function, $\mathcal{N}(\theta \mid \mu, \tau)$, 
into a standardized Gaussian probability density function, 
$\mathcal{N}(\tilde{\theta} \mid 0, 1)$ and the deterministic transformation, 
$\theta = \mu + \tau \cdot \tilde{\theta}$.  

Using
\begin{equation*}
\tilde{\boldsymbol{\theta}} = 
\left\{ \tilde{\theta}_{1}, \ldots, \tilde{\theta}_{N} \right\}
\end{equation*}
as parameters the latent Gaussian model can be specified by a product of independent 
probability density functions, 
\begin{align*}
\pi(\tilde{\boldsymbol{\theta}}, \mu, \tau)
&=
\pi(\tilde{\boldsymbol{\theta}} \mid \mu, \tau)
\cdot \pi(\mu, \tau)
\\
&=
\pi(\tilde{\boldsymbol{\theta}} )
\cdot \pi(\mu, \tau)
\\
&=
\prod_{n = 1}^{N} \mathcal{N} (\tilde{\theta}_{n} )
\cdot \pi(\mu, \tau).
\end{align*}

When incorporating individual observational models, however, the non-centered 
$\tilde{\theta}_{n}$ must be coupled to the population parameters in order to 
recreate each $\theta_{n}$,
\begin{align*}
\pi(\boldsymbol{y}, \tilde{\boldsymbol{\theta}}, \mu, \tau, \phi)
&=
\prod_{n = 1}^{N} \pi (y_{n} \mid \theta_{n} (\tilde{\theta}_{n}, \mu, \tau), \phi) \cdot
\mathcal{N} (\theta_{n} \mid \mu, \tau)
\cdot \pi(\mu, \tau)
\\
&=
\prod_{n = 1}^{N} \pi (y_{n} \mid \mu + \tau \cdot \tilde{\theta}_{n}, \phi) \, 
\mathcal{N} (\theta_{n} \mid \mu, \tau)
\cdot \pi(\mu, \tau).
\end{align*}

For weakly informative likelihood functions the posterior probability density 
function is dominated by the latent Gaussian probability density function, which 
now is free of the pathological funnel geometry.  On the other hand as the likelihood 
functions concentrate they strongly constrain the latent parameters, but only through 
the functions
\begin{equation*}
\mu + \tau \cdot \tilde{\theta}_{n}.
\end{equation*}
This constraint, however, induces its own funnel geometry!  In other words the 
non-centered parameterization yields a better geometry for weakly informative data 
and a worse geometry for strongly informative data, inverse to the behavior of the 
centered parameterization.

\subsection{Effective Metrics When Non-Centering}

\begin{figure*}
\centering
\subfigure[]{\includegraphics[width=2in]{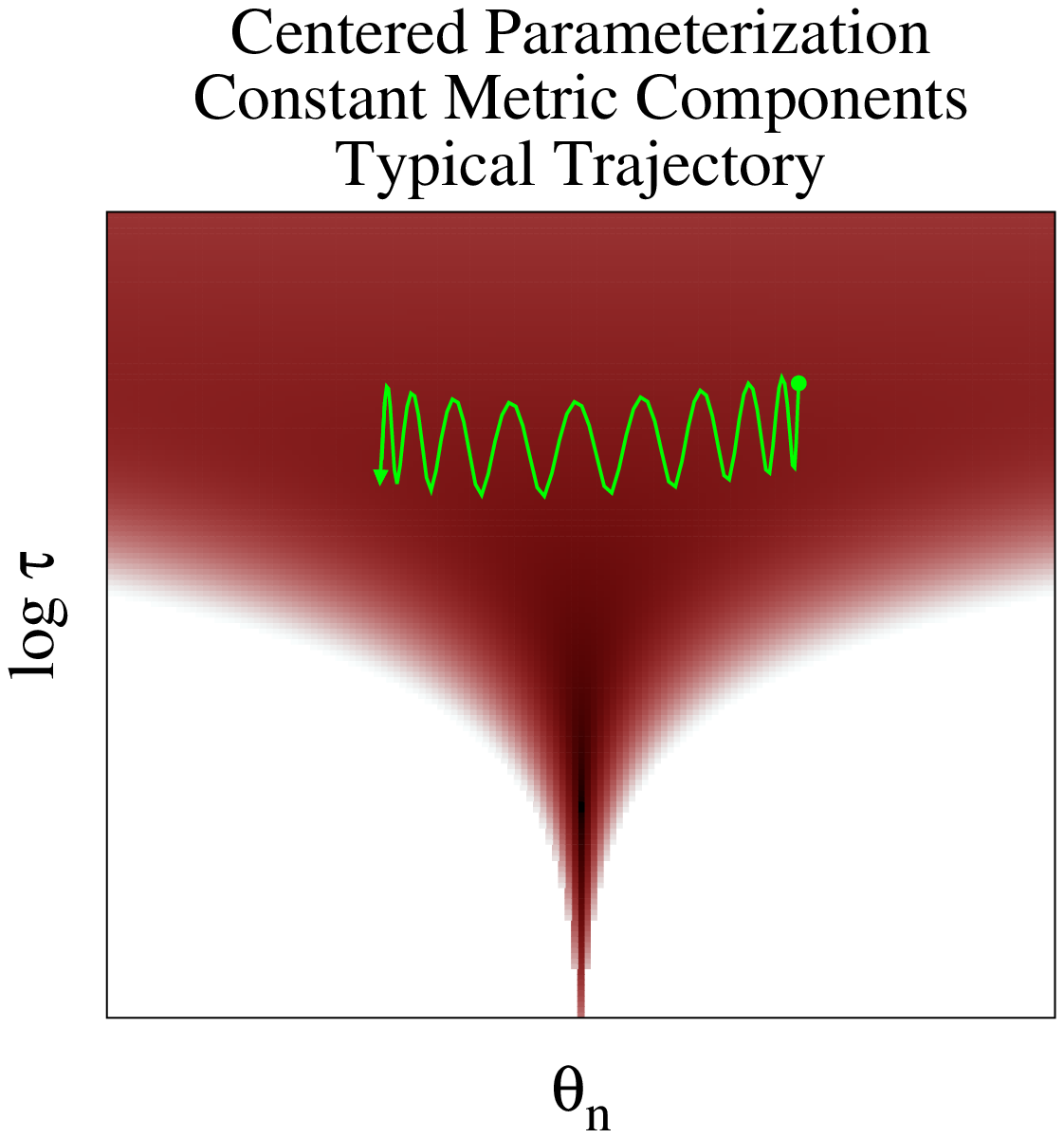}}
\subfigure[]{\includegraphics[width=2in]{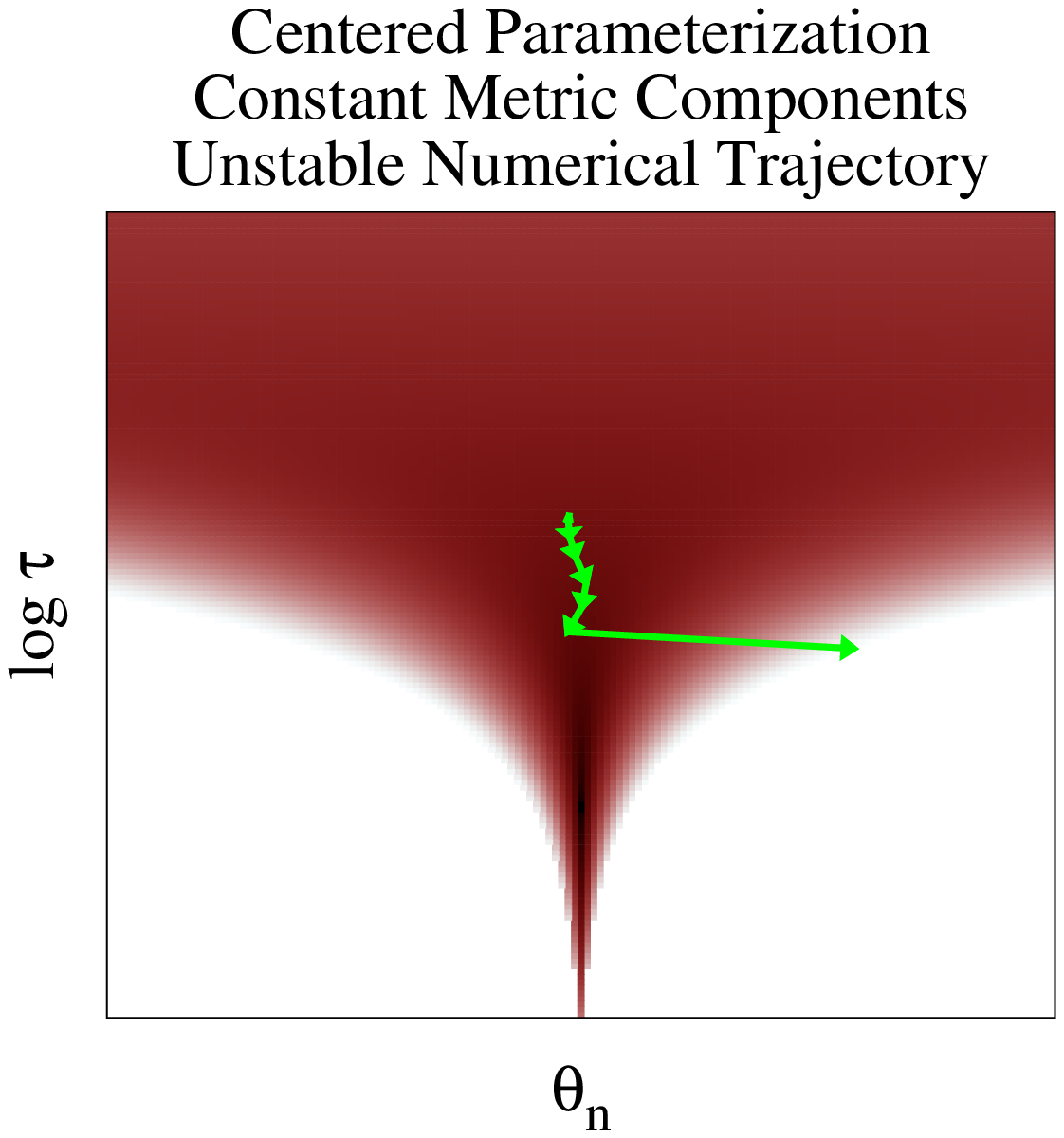}}
\caption{
A funnel geometry frustrates Hamiltonian Monte Carlo in numerous ways.
(a) Typical Hamiltonian trajectories span only a limited range of $\tau$ values 
and hence only slowly explore the entire distribution.  (b) Trajectories that
do penetrate deeper into the funnel are difficult to numerically integrate,
usually resulting in unstable numerical trajectories.
}
\label{fig:bad_trajectories} 
\end{figure*}

Riemannian algorithms that utilize constant metric components are not able to adapt 
to the rapidly varying curvature of the funnel and will consequently suffer when 
trying to explore posterior density functions corresponding to weakly-informed
likelihoods in the centered parameterization or strongly-informed likelihoods in 
the non-centered parameterization.  For example in Hamiltonian Monte Carlo this 
results in exact trajectories that tend to be restricted to narrow neighborhoods
of $\tau$, (Figure \ref{fig:bad_trajectories}a).  Moreover, when the trajectories
are lucky enough to venture deeper into the funnel their numerical integration 
becomes unstable (Figure \ref{fig:bad_trajectories}b).

\begin{figure*}
\centering
\includegraphics[width=2in]{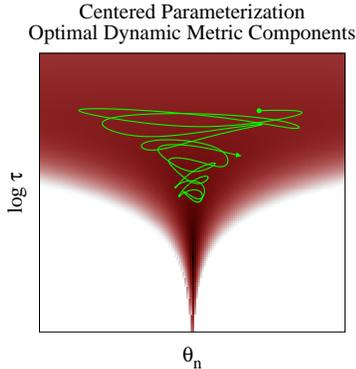}
\caption{
Using dynamic metric components that capture the second-order structure
of the funnel density function itself results in Hamiltonian trajectories
that span the entire funnel and explore much more efficiently.  The 
integration of these trajectories is more stable but also more difficult to
implement in practice.}
\label{fig:good_trajectory} 
\end{figure*}

One option around this pathology is to generalize the algorithms by allowing 
the metric components to vary and capture the Hessian structure of the posterior 
density function.  Although this results in exact and numerical trajectories 
that are much better behaved (Figure \ref{fig:good_trajectory}), the
general algorithms are significantly more challenging to implement.  In
Hamiltonian Monte Carlo, for example, this requires an implicit midpoint
symplectic integrator which needs a fixed point equation to be solved at
each iteration.

\begin{figure*}
\centering
\subfigure[]{\includegraphics[width=1.8in]{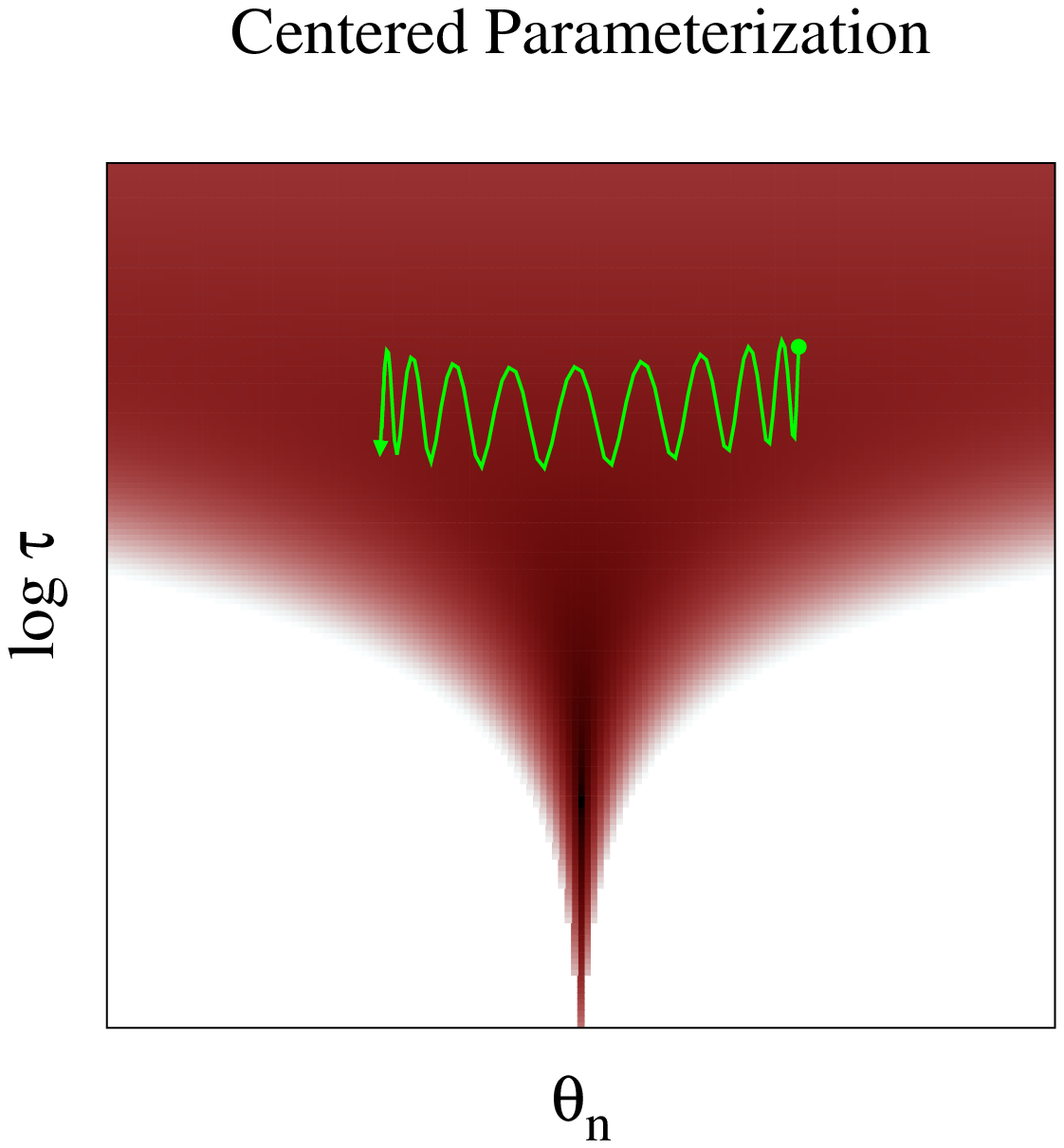}}
\subfigure[]{\includegraphics[width=1.8in]{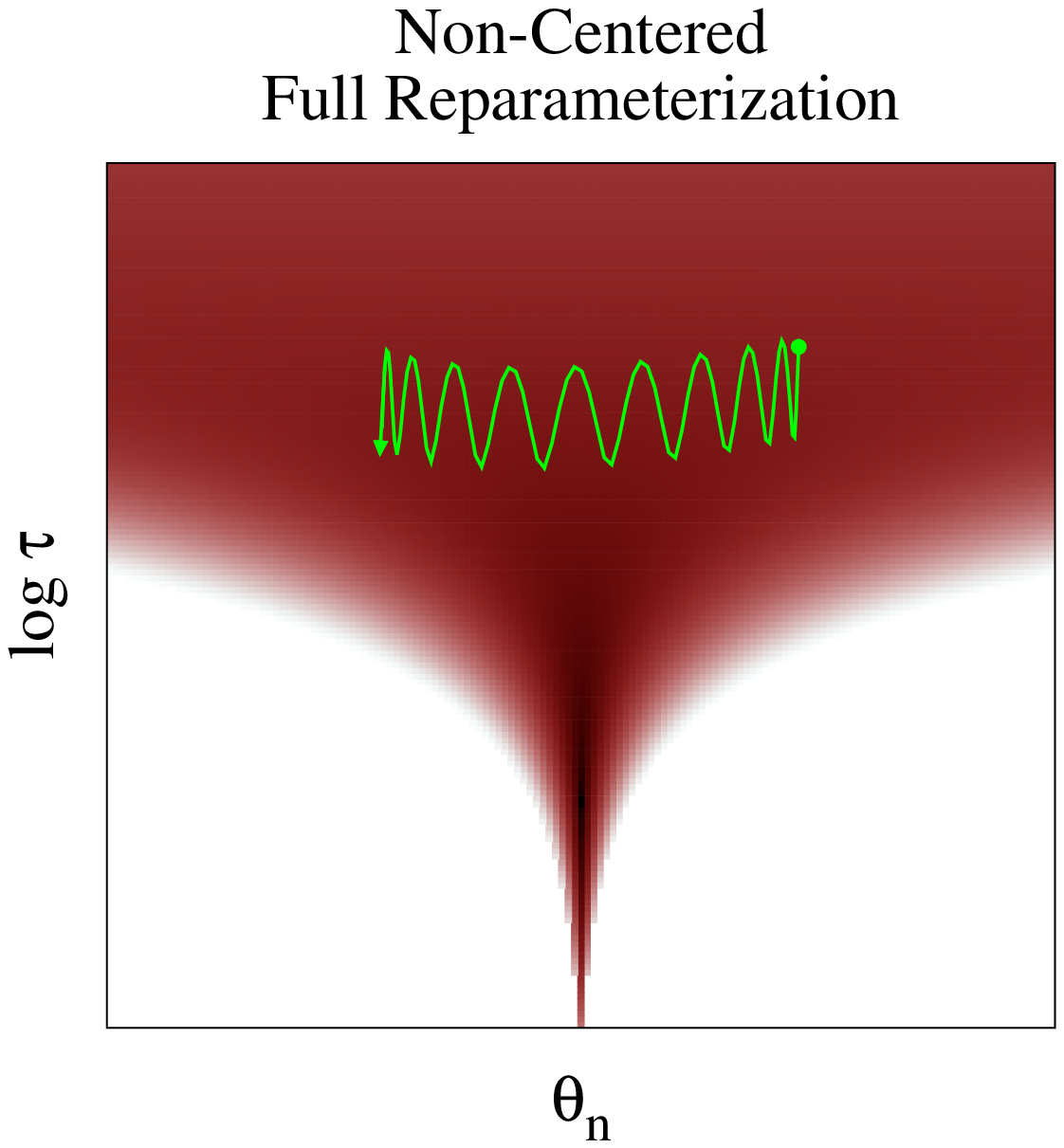}}
\subfigure[]{\includegraphics[width=1.8in]{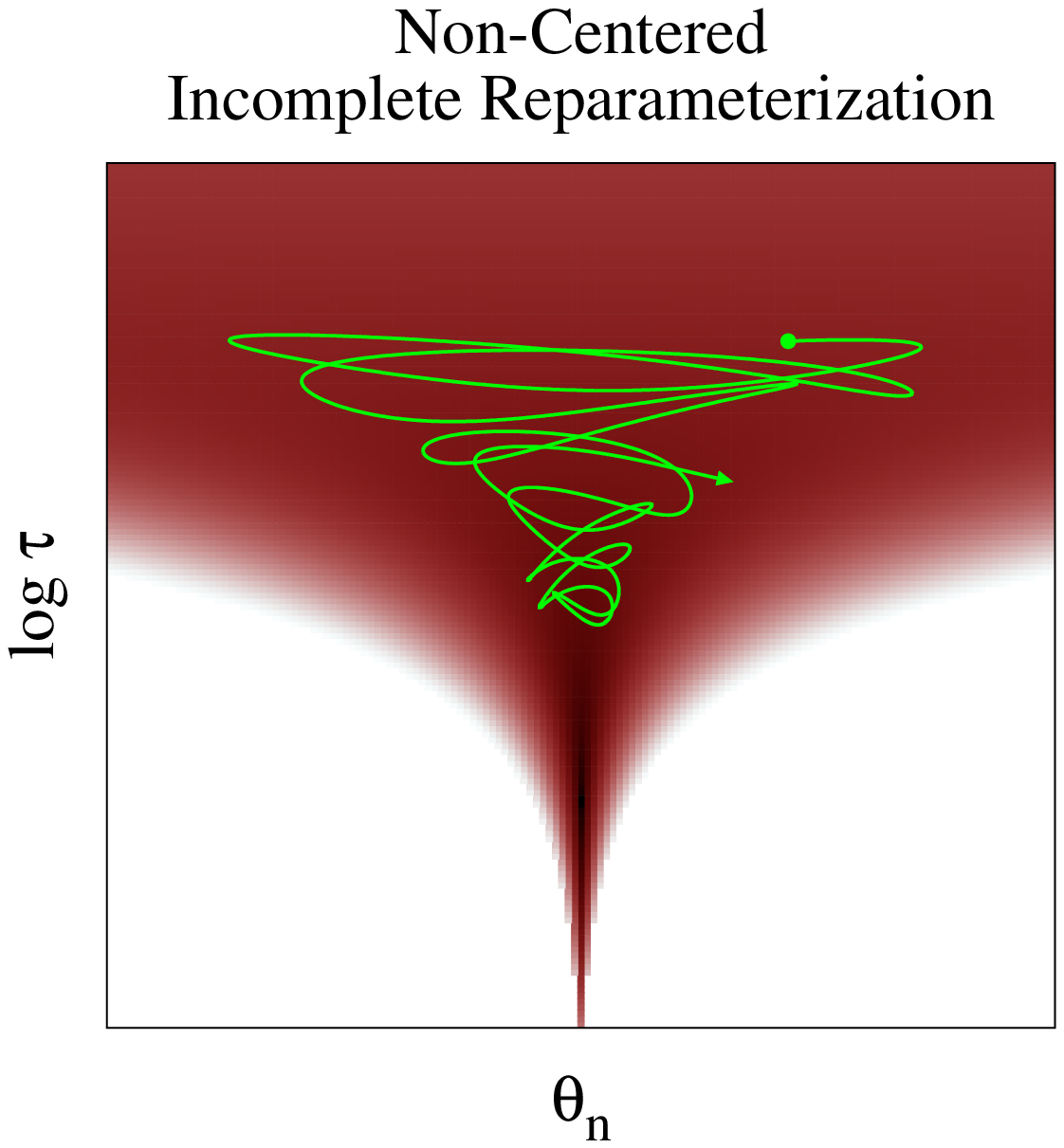}}
\caption{
Hamiltonian trajectories from a centered parameterization of a latent Gaussian model 
with only weakly-informative data are frustrated by the funnel geometry in the
posterior density function.  
(a) Typical trajectories explore only limited neighborhoods.  
(b) Applying a full reparameterization to a non-centered parameterization results in the 
same geometric system and hence the same Hamiltonian dynamics.  
(c) Applying an incomplete reparameterization to the non-centered parameterization, however, 
modifies the geometry, resulting in Hamiltonian dynamics better suited to explore the
funnel.
}
\label{fig:reparameterized_trajectories} 
\end{figure*}

As we learned in Section \ref{sec:equivalent_metric}, however, we can achieve the 
same behavior by applying a particular incomplete reparameterization
(Figure \ref{fig:reparameterized_trajectories}).  While we can't work out the 
ideal reparameterization analytically, we can investigate how well mapping 
between the canonical centered and non-centered parameterizations 
performs.

Consider having no observations so that the posterior distribution reduces to the
latent Gaussian model.  In this case empirical experience informs us that a centered 
parameterization will perform poorly, and we can achieve much better performance by 
transforming to a non-centered parameterization with the map 
\begin{align*}
\mu &= \mu
\\
\tau &= \tau
\\
\tilde{\theta} &= \frac{ \theta - \mu }{ \tau }.
\end{align*}
Because of the exchangeability of the $\theta_{n}$ we can analyze the efficacy of 
this reparameterization using any number of components.  To further simplify the
analysis let's consider only a single individual parameter, $\theta$.

If we fix the metric components to constants while applying this map to the parameters, 
exactly what effective metric do we induce?  To avoid any complications due to the 
positivity constraint on the population scale let's first reparameterize from $\tau$ 
to $\lambda = \log \tau$.  The non-centering transformation then becomes
\begin{equation*}
\tilde{\theta} = \frac{ \theta - \mu }{ e^{\lambda} }
\end{equation*}
with the Jacobian matrix
\begin{equation*}
J 
=
\frac{ \partial (\mu, \lambda, \tilde{\theta} ) }
{ \partial (\mu, \lambda, \theta) }
=
\begin{pmatrix}
1 & 0 & 1 \\
0 & 1 & 0 \\
- \cosh(\lambda) + \sinh(\lambda) & - e^{-\lambda} (\theta - \mu) & e^{-\lambda}
\end{pmatrix}
\end{equation*}
and determinant
\begin{equation*}
| J | = e^{-\lambda}.
\end{equation*}

If we assume that the initial metric components are equal to the identify matrix, 
with ones along the diagonal and zeroes elsewhere,
\begin{equation*}
g 
= 
\begin{pmatrix}
1 & 0 & 0 \\
0 & 1 & 0 \\
0 & 0 & 1
\end{pmatrix},
\end{equation*}
then the equivalent metric components are given by the matrix
\begin{align*}
g'(\theta, \mu, \tau)
&= 
J^{T}(\theta, \mu, \tau) \cdot g \cdot J(\theta, \mu, \tau)
\\
&=
\begin{pmatrix}
1 + e^{-2 \lambda} & e^{-2 \lambda} (\theta - \mu) & - e^{-2 \lambda} \\
e^{-2 \lambda} (\theta - \mu) & 1 + e^{-2 \lambda} (\theta - \mu)^{2} & - e^{-2 \lambda} (\theta - \mu) \\
- e^{-2 \lambda} & - e^{-2 \lambda} (\theta - \mu) & e^{-2 \lambda}
\end{pmatrix}
\\
&= \quad\quad\quad
\begin{pmatrix}
\;\;\,1 & \;\;\,0 & \;\;\,0 \\
\;\;\,0 & \;\;\,1 & \;\;\,0 \\
\;\;\,0 & \;\;\,0 & \;\;\,0 
\end{pmatrix}
+
e^{-2 \lambda} (\theta - \mu) \;\;
\begin{pmatrix}
\;\;\,0 & \;\;\,1 & \;\;\,0 \\
\;\;\,1 & \;\;\,0 & -1 \\
\;\;\,0 & -1 & \;\;\,0 
\end{pmatrix}
\\
& \quad +
e^{-2 \lambda}
\begin{pmatrix}
\;\;\,1 & \;\;\,0 & -1 \\
\;\;\,0 & \;\;\,0 & \;\;\,0 \\
-1 & \;\;\,0 & \;\;\,1 
\end{pmatrix}
+
e^{-2 \lambda} (\theta - \mu)^{2}
\begin{pmatrix}
\;\;\,0 & \;\;\,0 & \;\;\,0 \\
\;\;\,0 & \;\;\,1 & \;\;\,0 \\
\;\;\,0 & \;\;\,0 & \;\;\,0 
\end{pmatrix}.
\end{align*}

To consider optimality we need an explicit target density function.
For the latent Gaussian model that means specifying prior density 
functions for $\mu$ and $\tau$.  Here let's consider unit Gaussian 
probability density functions for both $\mu$ and $\lambda$, or 
equivalently a log Gaussian probability density function for $\tau$.
The log joint target probability density function is then
\begin{equation*}
\log \pi(\theta, \mu, \lambda)
=
- \frac{1}{2} \left( \frac{ \theta - \mu }{ e^{\lambda} } \right)^{2} - \lambda
- \frac{1}{2} \mu^{2} - \frac{1}{2} \lambda^{2} + \mathrm{const}.
\end{equation*}
We can now compute the covariate derivative of this log probability 
density function with respect to our induced metric analytically;
here I use \cite{Headrick:2015} to compute the covariant derivative
symbolically.  In this case we get an exact cancelation,
\begin{equation*}
\Delta(q) 
= g'(q) -  \nabla^{2} \log \pi(q)
=
\begin{pmatrix}
0 & 0 & 0 \\
0 & 0 & 0 \\
0 & 0 & 0
\end{pmatrix}.
\end{equation*}
The non-centering transformation is exactly the optimal incomplete reparameterization!  
Running a Riemannian algorithm with unit metric and a non-centered parameterization of 
the target distribution is equivalent to running an algorithm whose metric captures 
the local second-order differential structure of the latent Gaussian model but in a 
centered parameterization.

If the prior densities on $\mu$ and $\lambda$ have non-unit scales then we can 
maintain optimality by matching those scales in the diagonal elements of the initial 
metric components.  In particular an adaptive algorithm that sets the diagonal elements 
of the initial metric components to the variance of each parameter function will be able 
to sustain optimality for arbitrary prior scales.

Although we cannot in general extend this analytic analysis to nontrivial observational
models, we can use the geometric perspective to provide qualitative information about the
influence of those models.  For example this optimality criterion considers only the 
first-order and second-order partial derivatives of the target probability density function, 
which means that the influence of a nontrivial observational model is captured within the 
first-order and second-order behavior of the likelihood functions.  Contrast this to the
Fisher information matrix, which captures the same information but only in expectation 
with respect to possible observations.

At the same time the geometric analysis is useful for motivating even further questions.  
For example the common non-centering reparameterization is geometrically optimal only 
for the log Gaussian prior density on $\lambda = \log \tau$.  This prior choice, however, 
suppresses the limit $\tau \rightarrow 0$ corresponding to an identical, independently 
distributed ensemble of individuals.  In statistical modeling we typically want to include 
that homogeneous limit in the prior distribution and instead appeal to prior density 
functions that don't suppress zero such as half Gaussian probability density functions.  
This immediately raises the question of in what ways non-centering is suboptimal for these 
half-Gaussian priors and what practical consequences would that have for models where data 
cannot exclude those homogeneous configurations.

By isolating the interactions that influence the performance of Riemannian algorithms, 
the geometric perspective identify the features of our model that contribute to these
interactions and hence require the closest examination.

\section{Conclusion}

Placing inherently geometric algorithms like random walk Metropolis-Hastings, Langevin 
Monte Carlo, and Hamiltonian Monte Carlo into a proper geometric framework enables a
wide range of theoretical analyses.  In particular we can use the Riemannian structure 
of these algorithms to quantify the affect of incomplete reparameterizations.  We can
even motivate incomplete reparameterizations that optimize the local geometry for all 
of these algorithms at the same time.

Here we demonstrated this analysis on a particularly simple Gaussian latent model where 
we could analytically prove the geometric optimality induced by non-centering the 
natural parameterization, at least in the case of non-influential data.  Although the 
analytic results don't immediately generalize to more complex systems, the qualitative
insights stretches beyond the confines of that simple system.  They suggests important 
questions and connections that may ultimately lead to important insights in more general
circumstances.  It also suggests empirical studies, such as correlating the optimality
criterion $\left| \Delta(q) \right|$ with effective sample size per iteration or other
quantifications of Markov chain Monte Carlo performance.

Insights about geometric algorithms, like those considered in this paper, will continue 
to be most efficiently mined by using geometric analyses that directly perceive their
fundamental structures.

\section{Acknowledgements}

I thank Dan Simpson for critical discussions about reparameterizations and geometry
as well as Luiz Carvahlo and Charles Margossian for helpful comments on this manuscript.

\bibliography{incomplete_reparameterizations}
\bibliographystyle{imsart-nameyear}

\end{document}